\documentclass{mn2e}
\usepackage{epsfig,natbib2,natbibmnfix}
\usepackage{amsmath}
\usepackage{mathrsfs}
\usepackage{subfigure}
\usepackage{url}
\usepackage[dvips]{color}
\usepackage{multicol}

\newcommand{\be}{\begin{equation}}
\newcommand{\ee}{\end{equation}}

\def\ltsima{$\; \buildrel < \over \sim \;$}
\def\simlt{\lower.5ex\hbox{\ltsima}}
\def\gtsima{$\; \buildrel > \over \sim \;$}
\def\simgt{\lower.5ex\hbox{\gtsima}}

\def\del#1{{}}

\def\msun{{\,{\rm M}_\odot}}

\title[Feeding SMBHs through supersonic turbulence and
    ballistic accretion]
{Feeding SMBHs through supersonic turbulence and ballistic accretion}

\author[Alexander Hobbs, Sergei Nayakshin, Chris Power, Andrew King]
       {\parbox{18cm}{Alexander Hobbs$^{*}$, Sergei Nayakshin, Chris Power,
           Andrew King}\vspace{0.3cm}\\
\noindent Dept. of Physics \& Astronomy, University of Leicester, Leicester, LE1 7RH, UK}

\begin{document}

\maketitle

\begin{abstract}
It has long been recognised that the main obstacle to accretion of gas onto
supermassive black holes (SMBHs) is large specific angular momentum. It is
feared that the gas settles in a large scale disc, and that accretion would
then proceed too inefficiently to explain the masses of the observed SMBHs.

Here we point out that, while the mean angular momentum in the bulge is very
likely to be large, the deviations from the mean can also be
significant. Indeed, cosmological simulations show that velocity and angular
momentum fields of gas flows onto galaxies are very complex. Furthermore,
inside bulges the gas velocity distribution can be further randomised by the
velocity kicks due to feedback from star formation.  We perform hydrodynamical
simulations of gaseous rotating shells infalling onto an SMBH, attempting to
quantify the importance of velocity dispersion in the gas at relatively large
distances from the black hole. We implement this dispersion by means of a
supersonic turbulent velocity spectrum.  We find that, while in the purely
rotating case the circularisation process leads to efficient mixing of gas
with different angular momentum, resulting in a low accretion rate, the
inclusion of turbulence increases this accretion rate by up to several orders
of magnitude. We show that this can be understood based on the notion of
``ballistic" accretion, whereby dense filaments, created by convergent
turbulent flows, travel through the ambient gas largely unaffected by
hydrodynamical drag. This prevents the efficient gas mixing that was found in
the simulations without turbulence, and allows a fraction of gas to impact the
innermost boundary of the simulations directly. Using the ballistic
approximation, we derive a simple analytical formula that captures the
numerical results to within a factor of a few. Rescaling our results to
astrophysical bulges, we argue that this ``ballistic'' mode of accretion could
provide the SMBHs with a sufficient supply of fuel without the need to channel
the gas via large-scale discs or bars. We therefore argue that star formation
in bulges can be a strong catalyst for SMBH accretion.
\end{abstract}

\renewcommand{\thefootnote}{\fnsymbol{footnote}}
\footnotetext[1]{E-mail: {\tt alexander.hobbs@astro.le.ac.uk }}

\section{Introduction}

The growth of supermassive black holes (SMBHs) at the centre of galaxy bulges
is known to be correlated with observable properties of the host spheroid. The
best established correlations are between the SMBH mass, $M_{\rm bh}$, and the
velocity dispersion, $\sigma$, i.e., the $M_{\rm bh}-\sigma$ relation
\citep{Gebhardt00, Ferrarese00}, and the $M_{\rm bh}-M_{\rm bulge}$ relation
\citep{Magorrian98,Haering04}, where $M_{\rm bulge}$ is the mass of the
bulge. Understanding the formation and growth mechanisms of SMBHs is therefore
believed to be important in determining the evolution of the larger host
systems.

High luminosities of high redshift quasars at $z \sim 6$ suggest SMBH masses
of up to $\sim 10^9 \msun$ \citep{Kurk07}, implying accretion onto the black
holes close to or even above the Eddington limit \citep{KingPringle07} for up
to 1 Gyr. Such a strong inflow is most likely the result of major mergers
between galaxies, which have been shown via hydrodynamical simulations to
drive gas inward through gravitational tidal fields \citep{BarnesHernquist91}.
Simulations have also shown that galaxy mergers within large dark matter halos
at $z \sim 14$ may be able to drive gas in to the centre of the galaxy at a
sufficient rate to grow a seed BH of $10^4 \msun$ to an SMBH of $10^9 \msun$
in the required time i.e., by $z \sim 6$ \citep{Li07}.  Recent cosmological
simulations of dark matter halos that include gas physics and black hole
feedback processes are able to build up the required mass by $z = 6$ via
extended Eddington-limited accretion \citep[e.g.,][]{Sijacki09}.

However, an ab-initio treatment of all the relevant physical processes
(gravity, star formation, magneto-hydrodynamics, radiative feedback etc.) over
the full range of scales required is still beyond current numerical
capabilities. Therefore, little is known about the mechanism(s) that connect
the large-scale flow ($\sim$ kpc) to the small-scale accretion flow in the
immediate vicinity of the SMBH ($\simlt 1$ pc). Current cosmological
simulations require sub-resolution prescriptions to encapsulate the accretion
process \citep[e.g.,][]{SpringelEtal05, BoothSchaye09, Sijacki09}. These
models commonly use Bondi-Hoyle accretion \citep{HoyleLyttleton39, Bondi44,
  Bondi52}, corrected upwards by orders of magnitude when the resolution
limitations cause an under-prediction of the desired SMBH growth
rate. Recently, \cite{DeBuhrEtal09} have performed hydrodynamical simulations
of major galaxy mergers employing a sub-grid model for viscous angular momentum
transport to dictate the accretion rate, the results of which suggest that the Bondi-Hoyle prescription
actually {\em over-predicts} accretion onto the SMBH by up to two
orders of magnitude. The applicability of the Bondi-Hoyle solution to the SMBH
accretion rate is therefore not clear.


Looking at the problem from the other end, where dynamical and viscous
timescales are short compared with cosmological timescales, it is expected that
geometrically thin and optically thick accretion discs \citep{Shakura73} form
due to a non-negligible amount of angular momentum in the gas.  While it is
possible to study such discs at a far higher resolution and with more physics,
the approach is obviously limited by the fact that the disc is commonly
assumed to just exist \emph{a priori}. How these discs operate on $\sim$
pc scales is also not clear as the discs are susceptible to self-gravity and
hence may collapse into stars
\citep{Paczynski78,Kolykhalov80,CollinZahn99,Goodman03,NC05}.

There is thus a genuine gap between these two scales and approaches.  The
purpose of this paper is to try to bridge this gap to some degree with the
help of numerical hydrodynamical simulations of gas accreting onto a SMBH
immersed in a static, spherical bulge potential. In particular, we study the
`intermediate-scale' flow, from the inner 100 pc of a galactic bulge (just
below the resolution limit for some of the better-resolved cosmological
simulations) to the inner parsec, where the gas would be expected to settle
into a rotationally-supported feature. Instead of assuming the presence of the
disc, we allow it to form self-consistently from the larger scale flow.  For
simplicity of analysis of the results, we start with a spherically-symmetric
shell of gas of constant density, where the gas is isothermal throughout the
simulation. The first non-trivial element in our study is the rotation of the
shell, which should force the gas to settle into a rotating disc. 

While this turns out to be an interesting system in itself, forming a narrow ring
rather than a disc, we feel motivated to consider a more complex initial
velocity field in the shell. First of all, cosmological simulations show
the presence of cold streams \citep[e.g.,][]{DekelEtal09} immersed into a hotter,
lower density medium. These complex gaseous structures must also propagate to smaller
scales within the bulge. Secondly, the abundance of gas required to fuel the
SMBH should also fuel star formation in the host. Massive stars deposit large
amounts of energy and momentum in the surrounding gas \citep{Leitherer1992},
which is probably one of the dominant sources of turbulence in the
interstellar medium \citep[e.g.,][]{McKeeOstriker77,MacLowKlessen04}.  In
general, where we can observe it, it is evident that turbulence is ubiquitous
in astrophysical flows over a wide range of scales and systems \citep[for a
  review see e.g.,][]{ElmegreenScalo04}. Strong supersonic turbulence is a key
ingredient in all modern simulations of star formation
\citep[e.g.,][]{KrumholzEtal07,Bate09b}.

Therefore, we must expect that, far from the SMBH, while the specific angular
momentum of gas, $l$, is too large to be captured by the SMBH within the
central parsec (or less), the dispersion in $l$ is also large. It is not
unreasonable to expect that there will be some gas counter-rotating with
respect to the mean rotation of the flow. To test the importance of these
ideas for accretion of gas onto SMBHs, we draw on the machinery developed for 
turbulent flows in the star formation field by adding a turbulent
velocity spectrum to the gas in the initial shell. By varying the
normalisation of the turbulent velocities we essentially control the
dispersion in the initial gas velocities.

We find that the role of the turbulence and in general any disordered
supersonic velocity field is two-fold. First of all, it broadens the
distribution of specific angular momentum, setting some gas on lower angular
momentum orbits. Secondly, as is well known \citep[e.g.,][]{McKeeOstriker07},
turbulence creates convergent flows that compresses gas to high densities. We
find that the dynamics of such regions can be described reasonably well by the ballistic
approximation in which hydrodynamical drag and shocks are of minor importance
for the high density ``bullets'' moving through the lower density
background fluid.

These two effects increase the accretion rate onto the SMBH by up to several
orders of magnitude, largely alleviating the angular momentum barrier
problem. Note that several authors have already suggested that star formation
feedback and supersonic turbulence {\em inside} accretion discs \citep{WangEtal09} can promote SMBH
accretion by amplifying angular momentum transfer \citep{CollinZahn08,
  KawakatuWada08, ChenEtal09}.  An astrophysical conclusion from our
simulations is that star formation in the host can actually promote AGN
accretion.

The paper is arranged as follows. In Section 2 we outline the computational
method used, and in Section 3 we discuss our initial conditions. Section 4
presents an overview of the gas dynamics. Sections 5 \& 6 present the results
for the no turbulence and turbulent cases, respectively, and Section 7
details the fate of the gas that makes it inside the accretion radius. Sections 8
\& 9 comprise our discussion and conclusion respectively.

\section{Computational method}\label{sec:numerics}

We use the three-dimensional smoothed particle hydrodynamic (SPH)/N-body code
GADGET-3, an updated version of the code presented in
\cite{Springel05}. Smoothing lengths in the gas are fully adaptive down to a
minimum smoothing length of $2.8 \times 10^{-2}$ pc, which is much smaller
than the scales that we resolve in the simulation. The code employs the
conservative formulation of SPH as outlined in \cite{SpringelHernquist02},
with the smoothing lengths defined to ensure a fixed mass within the smoothing
kernel, for $N_{\rm neigh} = 40$. Each simulation below starts
with $N_{\rm SPH} \sim 4 \times 10^{6}$ particles, with an individual SPH
particle possessing a fixed mass of $m_{\rm sph} \approx 12 \msun$. The
Monaghan-Balsara form of artificial viscosity is employed \citep{MonaghanGingold83, Balsara95} with $\alpha = 1$ and $\beta = 2\alpha$.

The calculations are performed in a static isothermal potential with a central
constant density core to avoid divergence in the gravitational force at small
radii. We also include the black hole as a static Keplerian potential. The
mass enclosed within radius $r$ in this model is:
\begin{equation}
M(r) \; = \; M_{\rm bh} + \;
\begin{cases}
M_{\rm core} (r/r_{\rm core})^3 \;, \quad r < r_{\rm core}\\
M_a (r/a), \quad r \ge r_{\rm core}\;,
\end{cases}
\label{potential}
\end{equation}
where $M_{\rm bh} = 10^{8} \msun$, $M_a = 10^{10} \msun$, $a = 1$ kpc, and $r_{\rm core} = 20$ pc and
$M_{\rm core} = 2 \times 10^{8} \msun$. The one dimensional velocity dispersion of this
potential is $\sigma = (GM_a/2a)^{1/2} = 147$ km s$^{-1}$.

As we concentrate on the hydrodynamics of gas accreting on the SMBH, we make two
simplifying assumptions which avoid further non-trivial physics (to be
explored in our future work). First, the position of the SMBH is held fixed
during the simulations here. This is a reasonable approximation as all of our
initial conditions are either exactly spherically or azimuthally symmetric, or
have these symmetries when averaged over the entire simulation volume (e.g.,
turbulence is assumed to be isotropic in this case). Allowing the black hole to move
self-consistently would require relaxing the static potential assumption.

Secondly, gravitational forces between the particles are switched off to avoid
complications that might arise from gas self-gravity. While this could be
viewed as a shortcoming of our work, we believe that inclusion of self-gravity, ensuing star
formation and stellar feedback would only strengthen our results and
conclusions. The gravitational contraction of gas clouds would create even higher
density contrasts and thus make ballistic trajectories even more likely. Star
formation feedback would drive its own turbulence and hence amplify the
effects of the turbulence that we seed.

One further simplification is that the gas is isothermal throughout the
entirety of the simulations. This is a fair assumption as cooling times are
short for the high densities we are considering.  Accretion of gas onto the
SMBH is modelled with the ``accretion radius'' approach - particles that come
within an accretion radius of $r_{\rm acc} = 1$ pc are removed from the
simulation and we track the total mass and each component of the net angular
momentum vector of the accreted particles. This contrasts with the Bondi-Hoyle
accretion formulations that are frequently used in cosmological simulations
\citep[e.g.,][]{Sijacki09, BoothSchaye09}. We
believe the accretion radius approach, frequently used in the star formation
field \citep[e.g.,][]{Bate95}, is essential for the problem at hand as it
prevents unphysical accretion of SPH particles with too large an angular
momentum, whereas the Bondi-Hoyle formulation does not.

The units of length, mass, time and velocity used in the simulations are,
respectively, $r_{\rm u} = 3.08 \times 10^{21}$ cm (1 kpc), $M_{\rm u} = 2
\times 10^{43}$ g ($10^{10} \msun$), $t_{\rm u} = 1/\Omega(r_{\rm u}) \approx
5$ Myr, and $v_{\rm u} = (GM_a/a)^{1/2} = 208$ km s$^{-1}$.

\section{Initial conditions}\label{sec:ic}

The starting condition for all our simulations is that of a uniform density,
spherically symmetric gaseous shell centered on the black hole. The inner and
outer radii of the shell are $r_{\rm in} = 0.03$ and $r_{\rm out} = 0.1$ kpc,
respectively, for most of the simulations (see Table 1 in the Appendix). In
principle one can expect the outer radius of the shell to be much larger in a
realistic bulge with effective radius of a few kpc, but we are forced to limit
the dynamic range of the simulations due to computational
resources. Furthermore, we believe we understand how our results scale with
the outer radius of the shell (cf. Section \ref{sec:outerradius}), and hence
the dynamic range limitation does not actually influence our conclusions.

The total mass of the shell is $M_{\rm sh} = 5.1 \times 10^{7} \msun$.  To
minimise initial inhomogeneities we cut the shell from a relaxed, glass-like
configuration. The initial velocity field is composed of two parts: net
rotation and a seeded turbulent spectrum. The net rotation is described by the azimuthal velocity component, $v_\phi = v_{\rm rot}$,
where $v_{\rm rot}$ is a parameter of the simulation (see Table \ref{table1}
in the Appendix). In all our runs, $v_{\rm rot}$ is below the circular
velocity at the shell radii, meaning that our initial conditions are not in
equilibrium. We stress that this is deliberate, as our investigation here is
the formation of a disc from infalling gas, rather than from already
rotationally-supported gas.

Our approach to setting up the turbulent velocity field follows that of 
Dubinski, Narayan \& Phillips (1995). We assume a Kolmogorov power spectrum,
\begin{equation}
  \label{eq:pvk}
  P_v(k) \sim k^{-11/3},
\end{equation}
\noindent where $k$ is the wavenumber. The key assumption here is that the 
velocity field is homogeneous and incompressible, and so we can define
$\vec{v}$ in terms of a vector potential $\vec{A}$ such that 
$\vec{v}=\nabla \times \vec{A}$. This is useful because the components of $A$ 
can then be described by a Gaussian random field with an associated power 
spectrum 
\begin{equation}
  \label{eq:ak}
  P_A(k) \sim k^{-17/3}.
\end{equation}
\noindent This is a steep power-law, implying that the variance in 
$|\vec{A}|$ will diverge sharply as $k$ decreases, and so we follow Dubinski,
Narayan \& Phillips (1995) and introduce a small scale cut-off $k_{\rm min}$.
Equation~\ref{eq:ak} can then be written as
\begin{equation}
  \label{eq:ak}
  <|A_k|^2> = C (k^2+k_{\rm min}^2)^{-17/6},
\end{equation}
\noindent where $C$ is a constant that sets the normalisation of the
velocities. For our purposes we set it equal to unity and normalise the 
velocity field once the statistical realisation has been generated. Physically
the small scale cut-off $k_{\rm min}$ can be interpreted as the scale 
$R_{\rm max} \simeq k_{\rm min}^{-1}$ as the largest scale on which the 
turbulence is likely to be driven.

Our approach to generating the statistical realisation of the velocity field
is straightforward. First we sample the vector potential $\vec{A}$ in Fourier 
space, drawing the amplitudes of the components of $\vec{A_k}$ at each point 
$(k_x,k_y,k_z)$ from a Rayleigh distribution with a variance given by 
$<|A_k|^2>$ and assigning phase angles that are uniformly distributed between 
$0$ and $2\pi$. Second we take the curl of $\vec{A_k}$,
\begin{equation}
  \label{eq:vk}
  \vec{v_k} = i\vec{k} \times \vec{A_k}
\end{equation}
\noindent to obtain the Fourier components of the velocity field. Finally we 
take the Fourier transform of this to obtain the velocity field in real space.
We use a periodic cubic grid of dimension $256^3$ when generating the
statistical realisation of the velocity field and we use tricubic
interpolation to estimate the components of the velocity field at the position
of each SPH particle.

The initial parameters for each run are given in Table \ref{table1}
in the Appendix.

\section{Overview of gas dynamics and main results}\label{sec:exp}

Before we embark on a quantitative study of the simulation results, we present
several snapshots from the simulations that illustrate graphically the nature
of the gas flow. In particular, here we discuss the overall gas dynamics for
two simulations that typify the extremes of behavior that we find - S30 and
S35 (cf. Table \ref{table1}). Both simulations have an initial rotation
velocity $v_{\rm rot} = 0.3$ which results in a mean circularisation radius of
$r_{\rm circ} = 0.019$ (see \S \ref{sec:circularisation}). Simulation S30 has
no imposed turbulence. Simulation S35 has turbulence characterised by $v_{\rm turb} \equiv \langle
v_{\rm turblent} \rangle = 1$, implying that gas turbulent motions are of the
order of the velocity dispersion in the bulge.

\subsection{A shell with no initial turbulence}\label{sec:no_turbulence}

\begin{figure*}
\centerline{\psfig{file=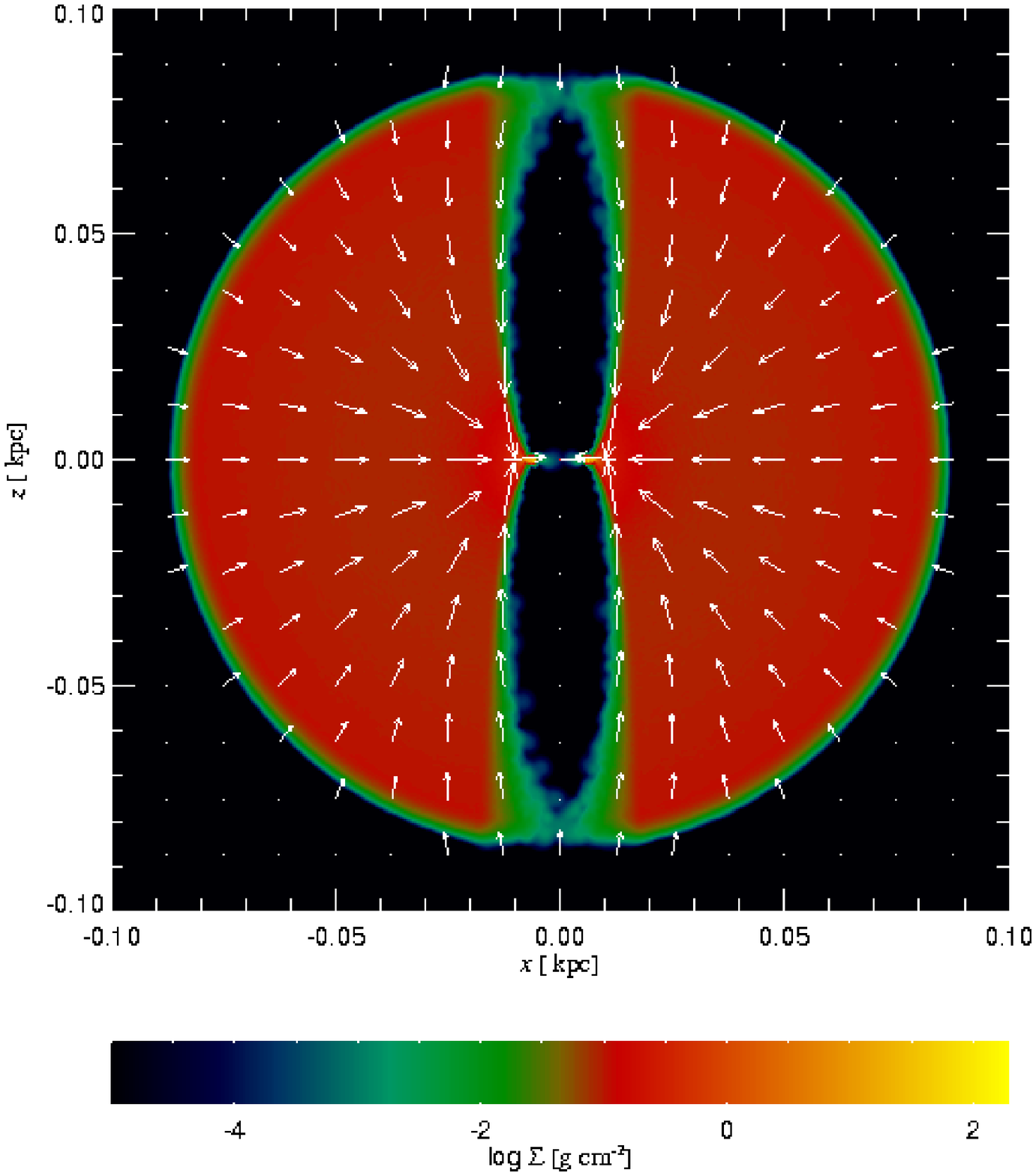,width=0.5\textwidth,angle=0}
\psfig{file=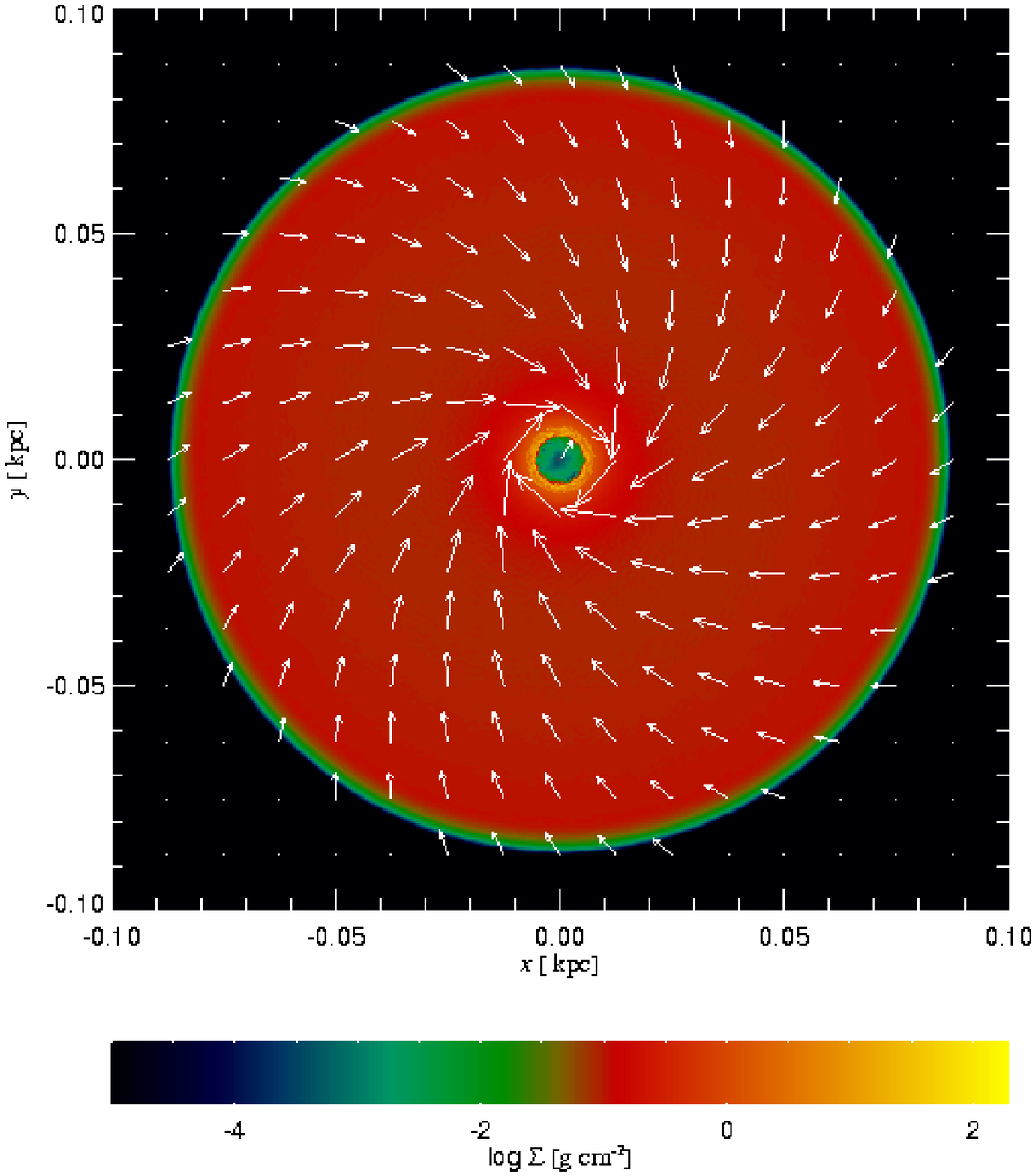,width=0.5\textwidth,angle=0}}
\caption{Edge-on angle-slice projection (see \S \ref{sec:no_turbulence}) of
  the gas flow at time $t=0.06$ in the simulation S30 (left) and top-down
  projection (right). The gas falls in on eccentric orbits, giving rise to a
  radial shock that propagates outwards as the disc forms.}
\label{fig:shell_t0.0_v0.3_030}
\end{figure*}

Our no turbulence initial condition acts as a base comparison for the
simulations with seeded turbulence. Figure \ref{fig:shell_t0.0_v0.3_030} shows the gas column density and the
velocity field in the angle-slice projection for the simulation S30 at
$t=0.06$. For projection along the $z$-axis, the gas
column density is calculated by
\begin{equation}
\Sigma(x, y) = \int_{-z(x,y)}^{z(x,y)} \rho(x, y) dz\;,
\label{sigma_proj}
\end{equation}
where the limits of the integration are given by $z(x,y)= r \tan \zeta$,
and $r = (x^2 + y^2)^{1/2}$. The angle $\zeta$ is chosen to be
$\tan \zeta = 1/5$ for Figures \ref{fig:shell_t0.0_v0.3_030} and \ref{fig:shell_t1.0_v0.3_030}. This projection method is convenient as
it permits an unobscured view into the central regions where the
black hole resides. 

Figure \ref{fig:shell_t0.0_v0.3_030} shows the gas flow at an early time, both
an edge-on (the left hand side panel of the figure) as well as a top-down
projection of the shell (the right panel). Due to a non-zero angular momentum,
gas in the shell is quickly pushed aside from the axis of symmetry, opening a
cylindrical cavity. Initially gas falls closer to the centre of the potential
than its circularisation radius, a consequence of highly eccentric orbits. As gas moves
inside $r_{\rm circ}$, centrifugal force exceeds gravitational
force. Interaction with neighboring gas streams results in ``radial''
shocks in the $xy$-plane.

A further set of shocks occurs due to gas initially moving supersonically in
the vertical direction.  As the gas from above the $z=0$ plane collides with
the gas falling in from below the plane, the particles are shocked and
accumulate at $z=0$ due to symmetry, forming a disc.  These two sets of shocks
mix gas with different angular momentum. We shall discuss this interesting
effect in greater detail below.

The overall dynamics of the simulation are thus relatively simple: angular
momentum conservation and symmetry dictate the formation of a geometrically
thin disc in the plane of symmetry of the shell. Accretion of gas onto the
SMBH would be expected to proceed in an accretion disc mode, if at all -- realistic models
show that star formation time scales are much shorter than viscous times,
depleting the disc of gas before it can feed the SMBH
\citep[e.g.,][]{Goodman03,NC05,NayakshinEtal07}.

We note that our particular choice of
velocity field, namely a constant azimuthal velocity, when combined with a spherical shell,
could be viewed as a somewhat peculiar initial condition. As we have
mentioned, the constant $v_\phi$ condition means that the polar regions of the shell are quickly
evacuated, the gas here spiralling outward in cylindrical radius to encounter gas
falling in with different angular momenta, and mixing with the latter. In fact
what we are modelling here is the simplest case of a flow that undergoes an \emph{angular momentum re-distribution shock}. 
We shall go into more detail on the consequences of this in Section
\ref{sec:laminar}, but for now we make the point that such a flow
is likely to occur when gas is accreting in a stochastic fashion from large
scales, perhaps as the result of a merger. The strong mixing of the gas with different angular momenta is exactly
the situation that we wish to explore here, and so we have implemented what is
essentially a symmetric case of this mixing process. We acknowledge that the spherical
setup is idealised, but it is also the best starting point from which to
embark on a laboratory of tests where the turbulence and rotation is
varied.

\begin{figure}
\centerline{\psfig{file=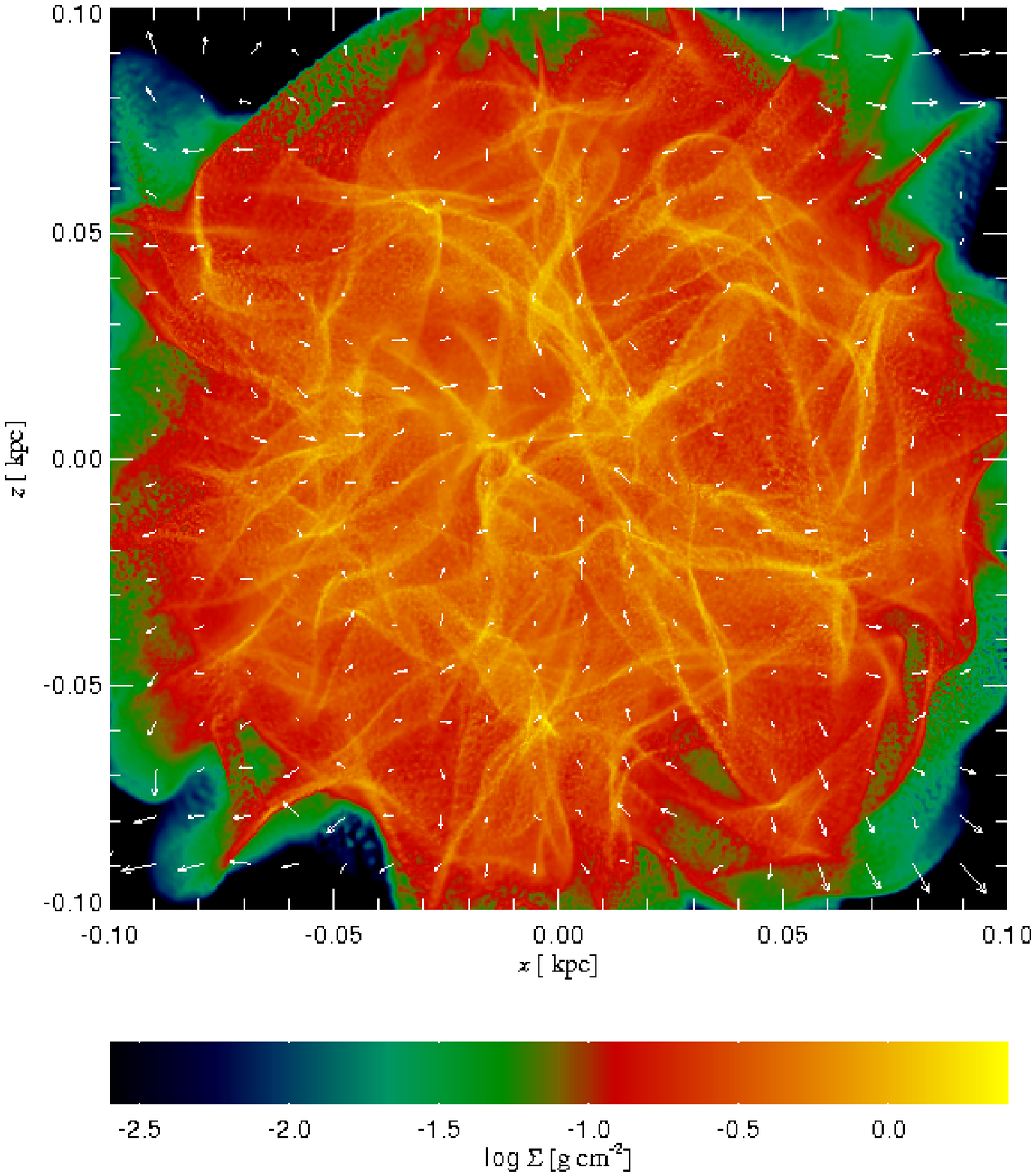,width=0.5\textwidth,angle=0}}
\caption{Projected gas column density in simulation S35 at time $t=0.02$,
  before gas has had a chance to accrete on the SMBH. Note the formation of multiple
  thin and dense filaments due to convergent turbulent velocity flows.}
\label{fig:s38_s10}
\end{figure}

\subsection{A shell with initial turbulence}\label{sec:dynamics_turbulent}

The flow of gas in this case is far more complicated than in
the case just considered. We first show in Figure \ref{fig:s38_s10} the {\em
  full} slice projection of the simulation volume, e.g., we use here $x(z,y)=
r_{\rm out} = 0.1$ (c.f. equation \ref{sigma_proj}). The most outstanding
feature of the figure are the long dense filaments that form due to convergent
turbulent velocity flows. The density contrast between the filaments and
surrounding gas is over two orders of magnitude by this time.

\begin{figure*}
\centerline{\psfig{file=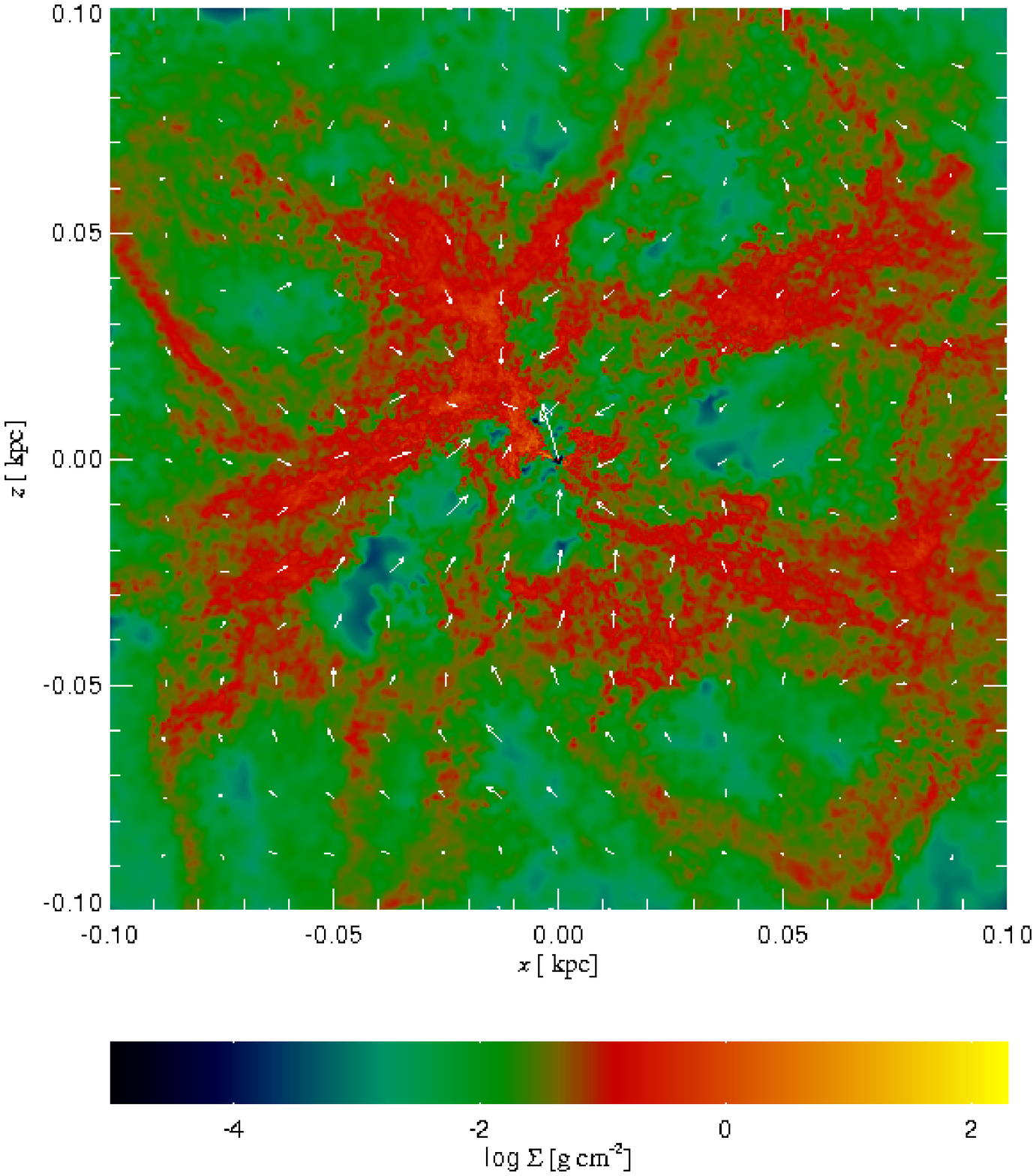,width=0.5\textwidth,angle=0}
\psfig{file=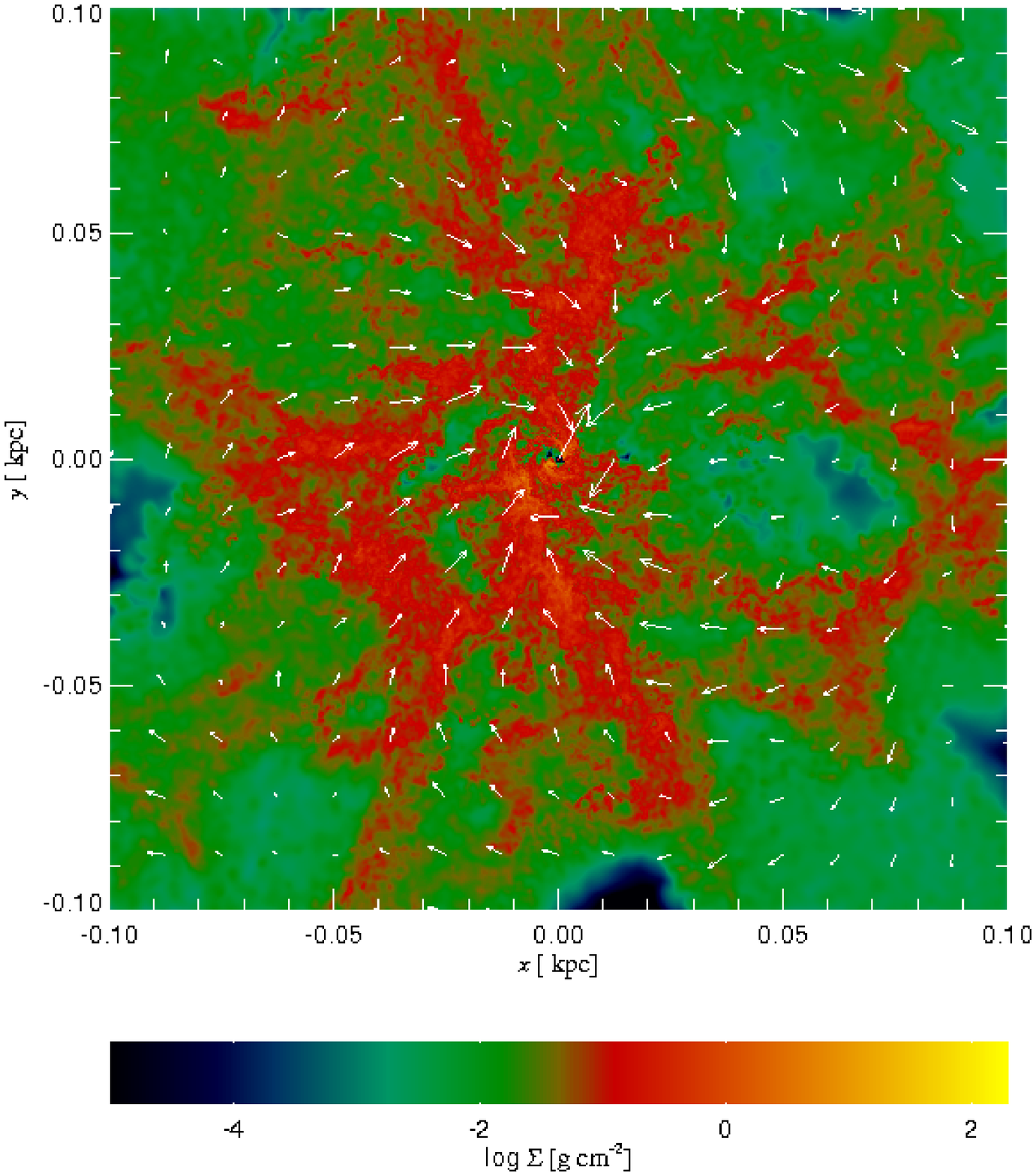,width=0.5\textwidth,angle=0}}
\caption{Edge-on angle-slice projection of the gas flow at time $t = 0.06$ in
  the simulation S35 (left) and top-down projection (right). The velocity
  field appears far more isotropic that in Figure
  \ref{fig:shell_t0.0_v0.3_030}, although an imprint of the imposed net
  rotation can still be seen.}
\label{fig:shell_t1.0_v0.3_030}
\end{figure*}

Figure \ref{fig:shell_t1.0_v0.3_030} shows both the y-axis and z-axis
projections of this simulation at the same time as \ref{fig:shell_t0.0_v0.3_030} did for the no initial turbulence case, $t=0.06$ .
Clearly, some of the filaments seen at the earlier time in Figure
\ref{fig:s38_s10} survive and actually penetrate into the innermost region
There are now density contrasts of as much as three orders of magnitude in 
regions that were completely uniform in simulation S30. We shall see that this
distinction is a crucial one for the dynamics of the gas and SMBH feeding.

\subsection{Turbulence and accretion}\label{sec:main_result}

\begin{figure}
\centerline{\psfig{file=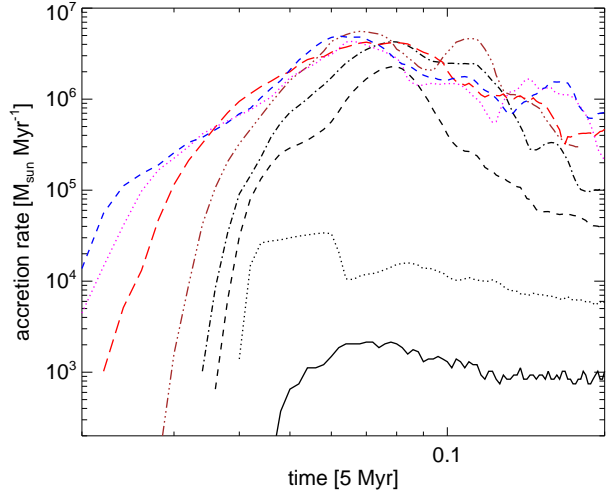,width=0.5\textwidth,angle=0}}
\caption{Accretion rate versus time for simulations S30--S37. This
  plot exemplifies the main result of the paper: the accretion rate on the
  black hole strongly increases with increasing levels of turbulence when
  rotation is present (see \S \ref{sec:main_result}).}
\label{fig:accreted_mass_v0.3}
\end{figure}

Figure \ref{fig:accreted_mass_v0.3} shows the mass accreted by the black hole
versus time in the simulations S30--S37, e.g., the same rotation velocity (and
thus angular momentum) but different levels of initial turbulence. This
demonstrates the main result of our study. The accretion rate onto the SMBH
strongly correlates with the strength of the imposed turbulence. The accretion
rate increases rapidly with increasing $v_{\rm turb}$ while $v_{\rm turb} \ll
v_{\rm rot}$, but then saturates at an approximately constant level for
$v_{\rm turb} \simgt v_{\rm rot}$.  The main qualitative explanation of the
simulations is that turbulence decreases the degree to which gas with
different angular momentum mixes, and creates gas streams with small angular
momentum. In particular,

\begin{itemize}
\item At low $v_{\rm turb} \ll v_{\rm rot}$, an increase in the turbulent
  velocity leads to greater variations in the density fluctuations that are
  created by the turbulent velocity flows before gas circularises (cf. Figure
  \ref{fig:s38_s10}). Greater density contrasts decrease the amount of angular
  momentum mixing, resulting in a disc rather than a narrow ring. The inner
  edge of the disc lies closer to the accretion radius of the simulations and
  hence feeds the SMBH more efficiently.

\item At $v_{\rm turb} \sim v_{\rm rot}$, random initial velocity fields set
  some gas on orbits with a vanishingly small angular momentum compared with
  the mean in the shell. The turbulent ``kick velocity'' in this case almost
  cancels the mean rotation for these regions. Since these regions move
  against the mean flow, they are also those that get strongly
  compressed. Reaching high densities, they continue to move on nearly
  ballistic trajectories, impacting the innermost region on randomly oriented
  orbits. Accretion in this regime is ``chaotic'' \citep{KingPringle07} rather
  than large-scale disc-dominated.

\end{itemize}

We shall spend the rest of the paper investigating these results in more
depth, suggesting and testing analytical explanations for the observed trends.

\section{Without turbulence: why ``laminar'' accretion is so low}\label{sec:laminar}

\subsection{Analytical estimates based on circularisation of gas}\label{sec:circularisation}

We shall now argue that the accretion rate of the simulation S30, the one with
rotation velocity $v_{\rm rot} = 0.3$ and no initial turbulence, is
surprisingly low compared with a straightforward and seemingly natural
theoretical estimate.

Let us start by estimating the fraction of gas that should be accreted by the
SMBH in our simulations.  Specific angular momentum of gas determines how
close to the SMBH it circularises. The angular momentum of a circular orbit is
given by $l = [G M(r) r]^{1/2}$. Using equation \ref{potential}, one can obtain a general solution for
the circularisation radius, $r_{\rm circ}$, for a given value of specific
angular momentum $l$. We give it in two extremes. If $r_{\rm circ} \ll r_{\rm
  core}$, the point-mass Keplerian value applies:
\begin{equation}
r_{\rm circ} = \frac{l^2}{GM_{\rm bh}}\;.
\label{rc_kepler}
\end{equation}
In the opposite case, $r_{\rm circ} > r_{\rm core}$, we have
\begin{equation}
r_{\rm circ} = \left[\frac{al^2}{GM_a} + \frac{M_{\rm bh}^2
    a^2}{4M_a^2}\right]^{1/2} - \frac{M_{\rm bh}
    a}{2M_a}\;,
\label{rc_iso}
\end{equation}
which simplifies to $r_{\rm circ} \approx l [a/GM_a]^{1/2}$ for $r_{\rm circ}
\gg r_{\rm core}$, when the second and the last terms on the right hand side
of the equation are small.

We now make the simplest possible assumption here by suggesting that gas
settles into a disc and that the distribution of gas in the disc follows the
distribution of gas over the circularisation radius initially, at time
$t=0$. Essentially, we assume that the $l_x$ and the $l_y$ components of the
initial angular momentum cancel out due to symmetry whereas the $l_z$
component is conserved without any exchange with neighboring cells. 

To estimate the fraction of gas that will end up inside the accretion radius
$r_{\rm acc} \ll r_{\rm core}$, we first note it is equation \ref{rc_kepler}
that should be used for the circularisation radius of gas with a given
$z$-projection of specific angular momentum, $l_z$. The requirement $l_z \leq
l_{\rm acc} = (GM_{\rm bh} r_{\rm acc})^{1/2}$ singles out a cylinder with
cross sectional radius of $r_1 = \sqrt{x^2 + y^2} < l_{\rm acc}/v_{\rm
  rot}$. The intersection of this cylinder with the shell $r_{\rm in} \le r
\le r_{\rm out}$ has volume $2 \pi r_1^2 (r_{\rm out} - r_{\rm in})$. For the
shell the fraction of the volume that can be accreted is then given by the
ratio of this volume to the total volume of the shell, $(4\pi/3) (r_{\rm
  out}^3 - r_{\rm in}^3) \sim (4\pi/3) r_{\rm out}^3$:
\begin{equation}
f_{\rm acc} \sim  \frac{3 GM_{\rm bh} r_{\rm acc} (r_{\rm out} - r_{\rm in})}{2 v_{\rm rot}^2 r_{\rm
    out}^3} \;.
\label{delta_vt0}
\end{equation}
For example, for $r_{\rm out} = 0.1$, $r_{\rm in} = 0.03$ and $v_{\rm rot} =
0.1$ this gives $f_{\rm acc} = 0.1$, and at $v_{\rm rot} = 0.3$ we have
$f_{\rm acc} \approx  10^{-2}$ (note that $G=1$ in code units).

The latter analytical estimate yields accreted mass $\eta_{\rm acc} M_{\rm
  shell} \approx 5\times 10^{5} \msun$, whereas Figure
\ref{fig:accreted_mass_v0.3} shows that the actually measured value to late
times is $\sim 10^3 \msun$. The analytical estimate thus significantly over
predicts the amount of accretion. 

\subsection{Shock mixing of gas: ring formation and end to accretion}

\begin{figure}
\begin{minipage}[b]{.48\textwidth}
\centerline{\psfig{file=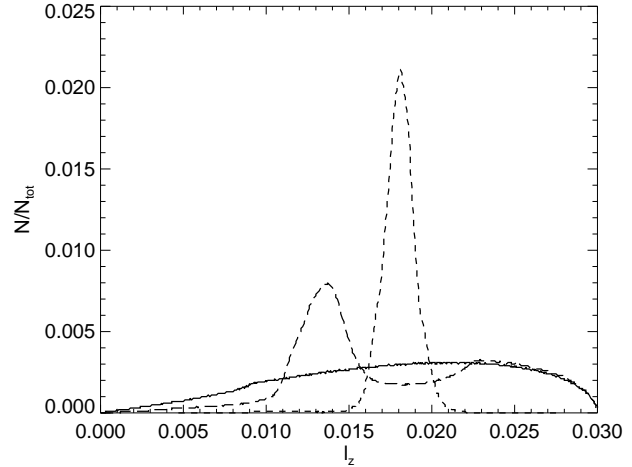,width=0.99\textwidth,angle=0}}
\end{minipage}
\caption{The distribution of gas in the simulation S30 over the $z$-component
  of the angular momentum at three different times: $t=0.0$, $t=0.1$, $t=0.2$
  for solid, long dashed and dashed curves respectively. Note how the angular momentum
distribution becomes narrower with time (through the action of shocks).}
\label{fig:lzgauss}
\end{figure}

Figure \ref{fig:lzgauss} explains why our simple analytical theory did not
work. Here we plot the distribution of gas in simulation S30 over the
$z$-component of the angular momentum vector of particles, $l_{\rm z}$, at
three different times. The initial distribution is spread over a broad range
of values, with a small but non neglibible fraction of gas having $l_z <
l_{\rm acc} \approx 0.003$, e.g. the angular momentum of the circular orbit at
$r_{\rm in}$. In our analytical estimate we assumed that this gas accretes
onto the black hole. However, this is not what happens.  The distribution of
angular momentum at the later time, $t=0.2$, shows a strong
radial mixing of gas with different angular momentum. The angular momentum
distribution narrows due to shocks and eventually becomes a highly peaked
Gaussian-like ring. 

Note that the deficit of gas in the innermost $r\simlt 0.01$ region in the
second curve in Figure \ref{fig:lzgauss} is caused not by the SMBH accretion
but by the shocks described above. Low angular momentum gas shocks and mixes
with high angular momentum material before it has a chance to travel into the
SMBH capture region, $r \le r_{\rm in}$. The mixing continues to late times
and the peak in the angular momentum distribution actually moves
outwards. This is to be expected as the gas that fell in earlier has a smaller
angular momentum and is in the path of eccentric orbits of gas falling in from
greater distances. 

We believe this strong mixing of different angular momentum orbits is quite a
general result of initially ``laminar'' flows.  Such flows thus initially form
rings rather than discs. While our simulations deliberately omit gas
self-gravity and hence star formation (\S \ref{sec:numerics}), previous
theoretical work shows that the fate of the material is then decided by
whether the viscous time of the ring is shorter than the star formation
consumption time scale. Most authors agree that large-scale gas discs form
stars more readily than they accrete \citep{Goodman03,NayakshinEtal07}. This
would suggests that ``laminar'' shells with a finite angular momentum would
form stars more readily than feed the SMBH.

\section{Accretion with seeded turbulence: why is it efficient?}

\subsection{Analytical expectations}\label{sec:turb_an}

\subsubsection{Weak turbulence}\label{sec:weak}

\begin{figure}
\begin{minipage}[b]{.48\textwidth}
\centerline{\psfig{file=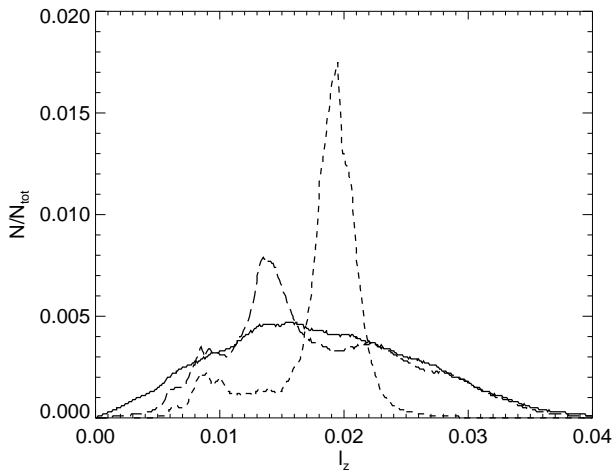,width=0.99\textwidth,angle=0}}
\end{minipage}
\caption{The same as Figure \ref{fig:lzgauss} but for simulation S31, with an initial
  turbulent velocity $v_{\rm turb} = 0.1$. There is a greater fraction of small angular
  material in this simulation initially, and more of it gets retained in
  the ``tail'' to small $l_z$, e.g., the inner disc, at late times.}
\label{fig:lzgauss_t0.1}
\end{figure}

\begin{figure}
\begin{minipage}[b]{.48\textwidth}
\centerline{\psfig{file=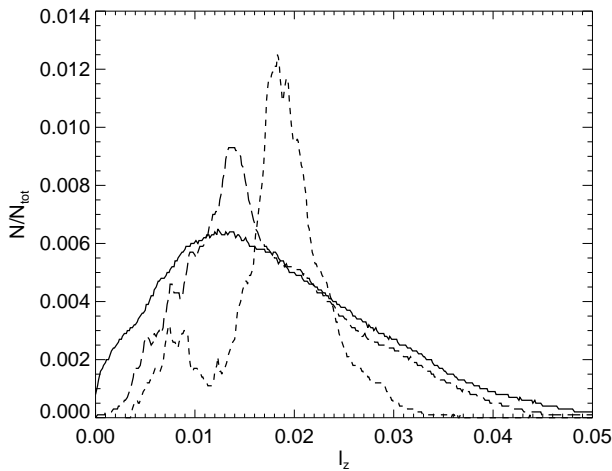,width=0.99\textwidth,angle=0}}
\end{minipage}
\caption{The same as Figure \ref{fig:lzgauss_t0.1} but for simulation S32. The features
noted in Figure \ref{fig:lzgauss_t0.1} are even more pronounced here.}
\label{fig:lzgauss_t0.2}
\end{figure}

First we consider the case of the turbulent velocity fields with $v_{\rm turb}
\ll v_{\rm rot}$.  Figures \ref{fig:lzgauss_t0.1} and \ref{fig:lzgauss_t0.2}
show the distribution of gas over the angular momentum for three different
times in the simulations S31 and S32, respectively. As in Figure
\ref{fig:lzgauss} for the no initial turbulence run S30, the initial angular
momentum distribution is broader than those at the later times. The small
angular momentum tail of the initial angular momentum distribution is more
pronounced for S31 and especially S32 compared with S30, which helps to
explain the higher accretion rates measured in these simulations.

The main effect, however, is the reduction in the shock mixing of gas. Consider a
``small'' angular momentum part of the distribution, to be definitive, at $l_z
\sim 0.01$. This part of the distribution completely disappears at later times
in simulation S30, being completely assimilated into the peak region. In
contrast, in simulations S31 and S32 this part of the distribution is only
reduced by a factor of 2-3 compared with the initial curves. 

Apparently, the initial turbulent velocity flows create density contrasts that
then propagate to smaller scales. The dynamics of a dense region are different
from that of a low density region, leading to a larger spread in the angular
momentum distribution at later times. This increases the amount of gas
captured by the inner boundary condition.

\subsubsection{Strong turbulence: ballistic accretion}\label{sec:vt_non_zero}

Now we consider the case when $v_{\rm turb} \simgt v_{\rm rot}$.  As we have
seen, the turbulent accretion rate is far larger in our simulations than that for
the ``initially laminar'' runs. We have also seen that differential and
chaotic velocity flows form strong density enhancements consistent with the
well known results from star formation studies
\citep[e.g.,][]{McKeeOstriker07}. High density regions could in principle
propagate through the mean density gas without much hydrodynamical drag, as
long as their column densities are much higher than that of the surrounding
gas. In this case we can approximate gas motion as ballistic and use the usual
loss cone argument \citep[e.g., page 406 in][]{ShapiroTeukolsky83} for black
hole accretion.

For a direct comparison to the simulation results, we consider a thin $\Delta r
\ll r$ shell of gas that has angular momentum $l \sim r v_{\rm rot}$, and
calculate the fraction of gaseous mass that has angular momentum small enough
to be captured within $r_{\rm acc}$. In doing so we assume that the energy and
angular momentum of gas are both conserved as the orbits are approximately
ballistic. The initial specific gas energy, $E$, is small compared with the
gravitational binding energy at $r\sim r_{\rm acc}$, therefore we can set
$E\approx 0$. An orbit that just reaches our inner boundary condition at
$r_{\rm acc}$ has radial velocity $v_{\rm r}(r_{\rm acc}) = 0$. This yields
the maximum specific angular momentum of gas that could still be accreted:
\begin{equation}
l_{\rm max}^{2} =  2 GM_{\rm bh}r_{\rm acc}\;.
\label{eq:lmax}
\end{equation}

The case of a shell that rotates slowly compared with the local circular
speed, i.e., $v_{\rm rot} \ll v_c = [GM_{\rm enc}(r)/r]^{1/2}$, is most
interesting, since the shell could not rotate faster than the circular speed
or else the centrifugal forces exceed gravity; if it rotates at the circular
speed then it is rotationally supported and would most likely form a disc
rather than a spherical shell. 

\subsubsection*{Assuming a monochromatic distribution}

As a starting point for this estimate, we approximate the
distribution of gas velocities in the shell with the turbulent velocity
spectrum by a monochromatic distribution $v = v_{\rm turb}$ randomly and
isotropically distributed in directions in the frame rotating with rotation
velocity $v_{\rm rot}$. We emphasise here that this is a
  simple assumption, but a sensible one, as on average one would expect the
  turbulent velocity field to have an isotropic character. We relate $v_{\rm turb}$ to the mean velocity
amplitude in the initial turbulent velocity distribution.

In the simplest case $v_{\rm rot} = 0$, the fraction of solid angle that
yields angular momentum smaller than $l_{\rm max}$ is approximately $l_{\rm
  max}^2/4(rv_{\rm turb})^2$. Therefore, the fraction of mass that could end
up inside the accretion radius is
\begin{equation}
\frac{\Delta M}{M_{\rm sh}} = \frac{GM_{\rm bh} r_{\rm acc}}{2 v_{\rm turb}^2
  r^2} \;,
\label{eq:dm_fraction}
\end{equation}
where $M_{\rm sh}$ is the mass of the shell. 

\begin{figure}
\begin{minipage}[h]{.48\textwidth}
\centerline{\psfig{file=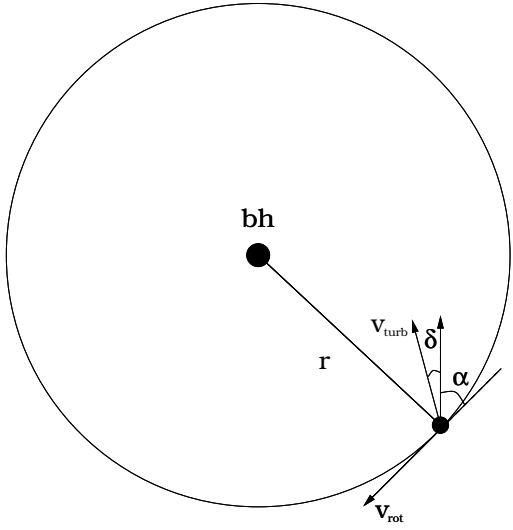,width=0.8\textwidth,angle=0}}
\end{minipage}
\caption{Schematic geometry of an isotropic (in the rest frame of the rotating
shell) wind, an analogy for the effect of turbulence on the loss-cone. The
shell rotates with velocity $v_{\rm rot}$ and the central axis of the
loss-cone is defined in terms of the angle $\alpha$.}
\label{fig:winddiagram}
\end{figure}

However, this approach neglects the rotation of the shell. For a non-zero rotation velocity
the loss-cone approach is valid only for $v_{\rm turb} \ge v_{\rm rot}$, and
equation \ref{eq:dm_fraction} must be modified to take account of the orbital
motion. Our derivation here is initially similar to \cite{Loeb04}, but taken
to second order when applying the small-angle approximation.

We start by considering the axis of zero angular momentum material for an
isotropic, monochromatic distribution at a point in orbit of the central black
hole. This is defined by
\begin{equation}
v_{\rm rot} - v_{\rm turb} \cos \alpha = 0
\label{eq:axiszeroL}
\end{equation}
where $\alpha$ is complementary to the angle between this axis and $r$,
as shown in Figure \ref{fig:winddiagram}. We consider a small perturbation in
angle about this axis, given by $\delta$, which defines the opening angle of
the loss-cone:
\begin{equation}
v_{\rm rot} - v_{\rm turb} \cos (\alpha \pm \delta) = \frac{l_{\rm max}}{r}
\label{eq:axisL}
\end{equation}
We expand this to second-order in the limit that $\delta \ll \alpha$, giving
us
\begin{equation}
v_{\rm turb} \frac{\delta^2}{2} \cos \alpha + v_{\rm turb} \hspace{0.05cm} \delta \sin \alpha
= \frac{l_{\rm max}}{r}
\label{eq:axisL_smallangle}
\end{equation}
where $\cos \alpha = v_{\rm rot}/v_{\rm turb}$. We therefore find for the
loss-cone opening angle the expression:
\begin{equation}
\delta = \left(\frac{2
  l_{\rm max}}{r v_{\rm rot}} + \frac{v_{\rm turb}^2 \sin^2 \alpha}{v_{\rm rot}^2}\right)^{1/2} - \frac{v_{\rm turb} \sin
  \alpha}{v_{\rm rot}}
\label{eq:delta}
\end{equation}
For an isotropic distribution in the rest frame of the orbiting body, the
fraction of the solid angle that will be accreted can be calculated via
\begin{equation}
\Omega_{\rm lc} = \frac{\Delta \Omega}{\Omega} \approx \frac{\pi \delta^2}{4 \pi}
\end{equation}
In this case, at a given $v_{\rm rot}$, our second-order approximation for the
loss-cone solid angle is:
\begin{equation}
\Omega_{\rm lc} = \left[\left(\frac{l_{\rm max}}{2 r v_{\rm rot}} + \xi \right)^{1/2} - \xi^{1/2} \right]^2
\label{eq:dm_fraction_moving}
\end{equation}
\noindent where $\xi = (v_{\rm turb}^2-v_{\rm rot}^2)/4 v_{\rm
      rot}^2$. 
We can express this in the limit of two extremes, depending on the relative
strength of the turbulence with respect to the rotation:

\vspace{0.07in}

\noindent i) the case when $v_{\rm
  turb} \gg v_{\rm rot}$:
\begin{equation}
\Omega_{\rm lc} \approx \left[\left(\frac{l_{\rm max}}{2 r v_{\rm rot}} + \frac{v_{\rm turb}^2}{4 v_{\rm
      rot}^2}\right)^{1/2} - \frac{v_{\rm turb}}{2 v_{\rm rot}} \right]^2
\label{eq:dm_fraction_gg}
\end{equation}
\noindent ii) when $v_{\rm turb} \approx v_{\rm rot}$:
\begin{equation}
\Omega_{\rm lc} \approx \frac{l_{\rm max}}{2 r v_{\rm rot}}
\label{eq:dm_fraction_equal}
\end{equation}

\subsubsection*{Assuming a Maxwellian distribution}

The treatment we have presented so far is not the whole picture, as we have assumed a
monochromatic distribution for the turbulence. In reality this is not the
case. The turbulent velocity distribution, over all three normally distributed
components, has a Maxwellian profile, both in the initial condition and
remaining approximately so in the high
density regions that form. We find that the distribution for the ballistic
approximation can therefore be fitted (to within $\sim 10\%$ error) by the expression:
\begin{equation}
f(v) = \frac{4}{v_{\rm turb}^3 \pi^{1/2}} v^2 e^{-(v/v_{\rm turb})^2}
\label{eq:fit}
\end{equation}
To obtain the total mass within the loss-cone we therefore need to integrate
this expression over the relevant solid angle:
\begin{equation}
M_{\rm lc} = \int_{0}^{\infty} \Omega_{\rm lc} f(v) dv
\end{equation}
were we should use equation \ref{eq:dm_fraction_moving} for the loss-cone
solid angle, but replacing $\xi$ with $\xi_{\rm v} = (v^2-v_{\rm rot}^2)/4 v_{\rm
      rot}^2$.  Unfortunately, performing the integration in this case is non-trivial
and must be done numerically. Luckily, we can obtain
a good analytical approximation by noting that the modified equation \ref{eq:dm_fraction_moving} is a maximum when
$v = v_{\rm rot}$, and falls off sharply with increasing $v$. The
majority of the accretion will therefore occur in the region of the distribution where
$v \approx v_{\rm rot}$ and so we can use equation \ref{eq:dm_fraction_equal}
in the integral instead, giving us:
\begin{equation}
 M_{\rm lc} = \int_{v_{\rm rot} - \Delta v}^{v_{\rm rot} + \Delta v} \frac{2
   l_{\rm max}}{r v_{\rm rot}v_{\rm turb}^3 \pi^{1/2}} v^2 e^{-(v/v_{\rm turb})^2} dv
\label{eq:integralapprox}
\end{equation}
where our limits are a perturbation either side of $v_{\rm rot}$, the size of
which we choose to be $\Delta v = v_{\rm rot}/4$ \footnote{it should be noted
  that this choice is somewhat arbitrary but our derivation here is of course approximate}. Performing the integral yields
\begin{equation}
M_{\rm lc} = \frac{l_{\rm max}}{2 r v_{\rm rot}} \cdot \psi + \frac{3 l_{\rm
    max}}{4 r v_{\rm turb} \pi^{1/2}} \cdot \varphi
\label{eq:mlc}
\end{equation}
where
\begin{displaymath}
\psi \equiv \text{erf}\left(\frac{5 v_{\rm rot}}{4 v_{\rm turb}}\right) -
  \text{erf}\left(\frac{3 v_{\rm rot}}{4 v_{\rm turb}}\right)
\end{displaymath}
\begin{displaymath}
\varphi \equiv e^{-9 v_{\rm rot}^2 /16
    v_{\rm turb}^2} - e^{-25 v_{\rm rot}^2 /16 v_{\rm turb}^2}
\end{displaymath}
For a thick shell we should generalise equation
\ref{eq:mlc} by using the density weighted average of $r$ over
the shell, $\langle r \rangle$. For an initially constant density profile used in our
simulations and $r_{\rm out} \approx 3 r_{\rm in}$, we have $\langle
r \rangle \approx 2 r_{\rm out}/3$, and hence
\begin{equation}
\frac{\Delta M}{M_{\rm shell}} = \frac{3 (GM_{\rm bh} r_{\rm acc})^{1/2}}{2 \sqrt{2} r_{\rm out} v_{\rm rot}} \cdot
\psi + \frac{9 (GM_{\rm bh} r_{\rm acc})^{1/2}}{4 \sqrt{2} r v_{\rm turb} \pi^{1/2}} \cdot \varphi
\label{eq:mlc_shell}
\end{equation}
As before, we can analyse this equation in various limits:

\vspace{0.07in}

\noindent i) when $v_{\rm turb} \gg v_{\rm rot}$:
\begin{equation}
\frac{\Delta M}{M_{\rm shell}} \approx \frac{(2 GM_{\rm bh} r_{\rm acc})^{1/2}}{r_{\rm out}
  v_{\rm turb}} \left(1 + \frac{v_{\rm rot}^2}{v_{\rm turb}^2}\right)
\label{eq:mlc_gg}
\end{equation}
\noindent ii) when $v_{\rm turb} \approx v_{\rm rot}$:
\begin{equation}
\frac{\Delta M}{M_{\rm shell}} \approx \frac{(GM_{\rm bh} r_{\rm
    acc})^{1/2}}{2 r_{\rm out}
  v_{\rm rot}}
\label{eq:mlc_equal}
\end{equation}
\noindent iii) when $v_{\rm turb} \ll v_{\rm rot}$:
\begin{equation}
\frac{\Delta M}{M_{\rm shell}} \approx \frac{9 (GM_{\rm bh} r_{\rm
    acc})^{1/2}}{4 \sqrt{2} r_{\rm out}
  v_{\rm turb} \pi^{1/2}} \cdot \varphi
\label{eq:mlc_ll}
\end{equation}
where we have dropped numerical factors of approximately unity.

\subsection{Detailed analysis of results}\label{sec:analysis}

Our ballistic approximation to the accretion rate yielded a result that
depends on several parameters of the simulations, e.g., the ``accretion''
radius $r_{\rm acc}$, the turbulent velocity $v_{\rm turb}$, the rotation
velocity $v_{\rm rot}$ and the dimensions of the shell. We shall now look at simulations covering a range in
the space of these parameters to discuss specific trends and determine whether the simulations support our
simple analytical theory.

\subsubsection{No net angular momentum case}

\begin{figure*}
\begin{minipage}[h]{.48\textwidth}
\centerline{\psfig{file=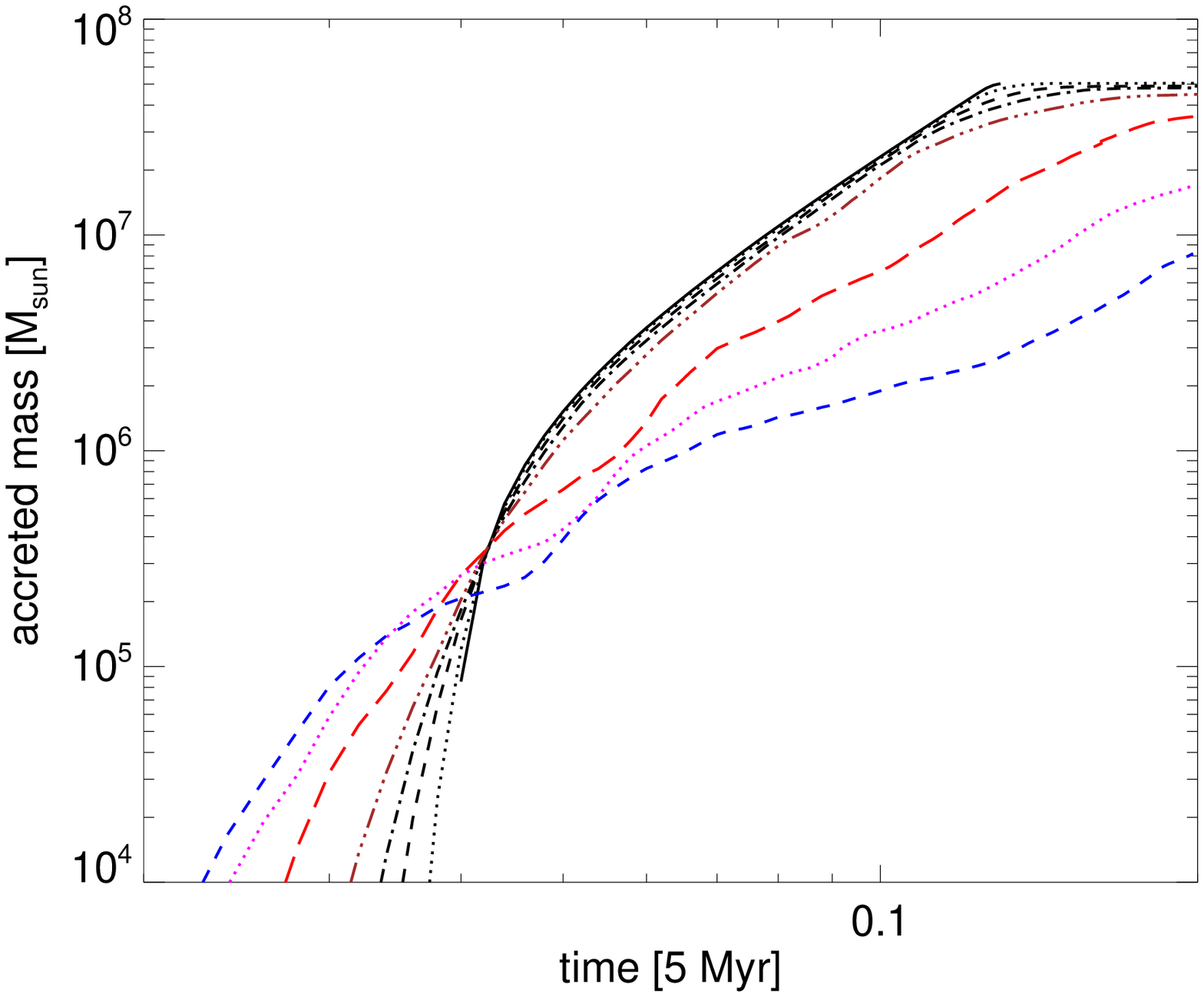,width=0.99\textwidth,angle=0}}
\end{minipage}
\begin{minipage}[h]{.48\textwidth}
\centerline{\psfig{file=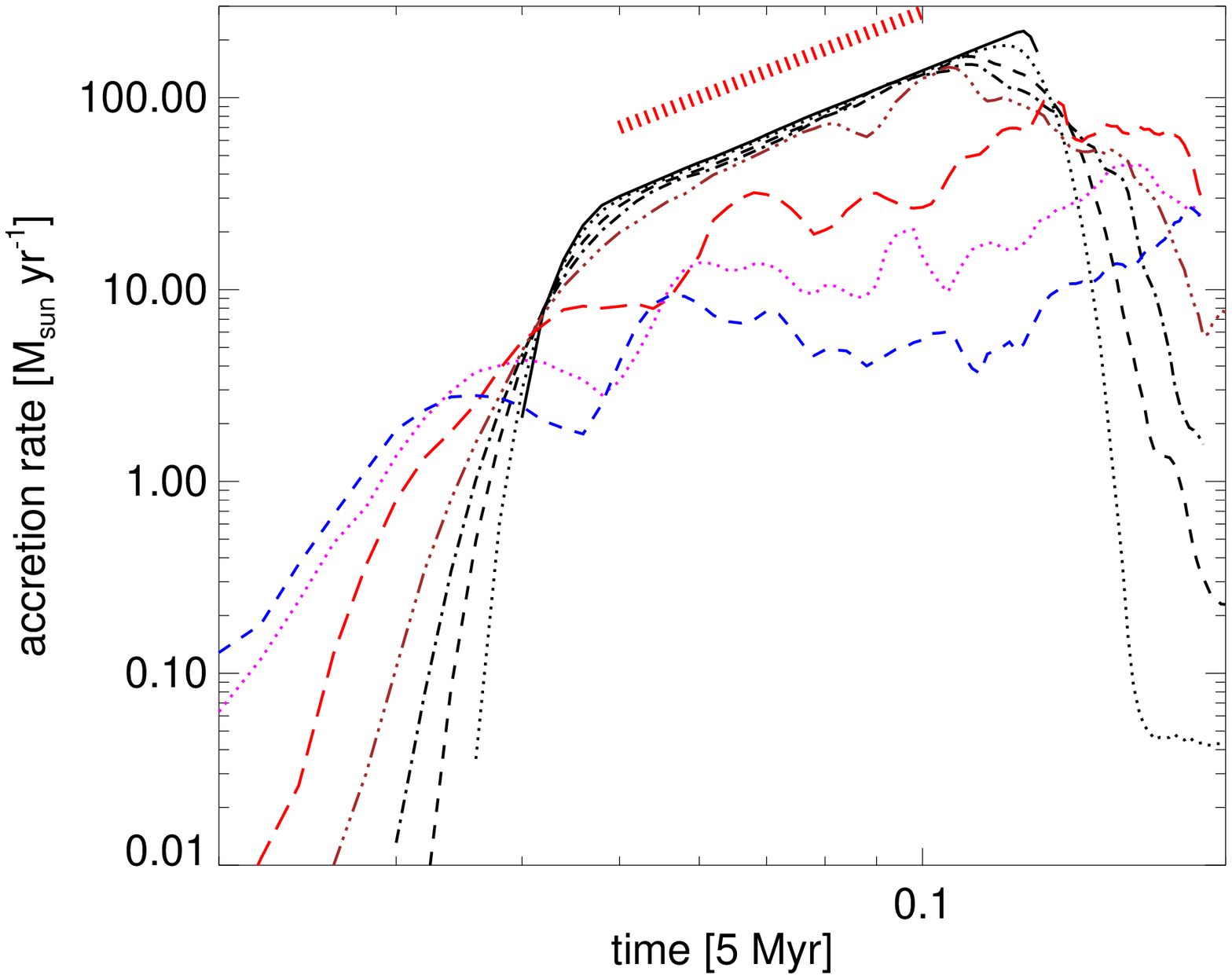,width=0.99\textwidth,angle=0}}
\end{minipage}
\caption{Accreted mass vs. time (left) and accretion rate vs. time (right) for
  a velocity field that contains no net rotation ($v_{\rm rot} = 0$). The
  different lines represent different strengths of the mean turbulent
  velocity, $v_{\rm turb}$: 0 (black solid), 0.1 (black dotted), 0.2 (black
  dashed), 0.3 (black dot-dashed), 0.5 (brown dot-dot-dash), 1.0 (red dashed),
  1.5 (pink dotted), and 2.0 (blue dashed). Analytical line (thick red dashed)
displays a $\dot M \propto t^2$ slope.}
\label{fig:vrot0}
\end{figure*}

Figure \ref{fig:vrot0} shows the accreted mass (left panel) and
the accretion rate (right panel) as functions of time for simulations S00 --
S07 (see Table 1). The rotation velocity for these runs is set to zero, and
the turbulent velocity parameter $v_t$ is varied from 0 (solid curve) to 2
(blue short dashed).

The situation is rather simple for these runs. For turbulent velocity
parameters much smaller than unity, the flow hardly deviates from that of a
freely falling shell with a negligible pressure. Due to the constant density
profile in our initial shell, this naturally leads to the accretion rate
increasing as $\dot M \propto t^2$ (see the dotted power law in the right
panel of Figure \ref{fig:vrot0}) starting from the free fall time of the inner
shell to that of the outer shell. As turbulence increases to $v_{\rm turb}
\simgt 1$, however, two effects are noteworthy. Firstly, the accretion
episode starts earlier and also finishes later. This is naturally due to the
spread in the initial radial velocities of the turbulent gas, with some
regions starting to fall in with negative velocities, and hence arriving at
the SMBH earlier, and others doing the opposite. There is also a significant
reduction in the accretion rate at the highest values of turbulence. We
believe this is due to the same effects already discussed in \S
\ref{sec:main_result}. Convergent turbulent flows create high density regions
with a large range of angular momentum. These regions do not mix as
readily as mean density regions and thus retain their angular momentum for
longer. This reduces the amount of gas accreted. However we note that in this
rather unrealistic set up (zero net angular momentum) gas accretion is still
very efficient, with $\sim 20 - 100$\% of the shell being accreted by time
$t=0.2$.

\begin{figure*}
\begin{minipage}[h]{.48\textwidth}
\centerline{\psfig{file=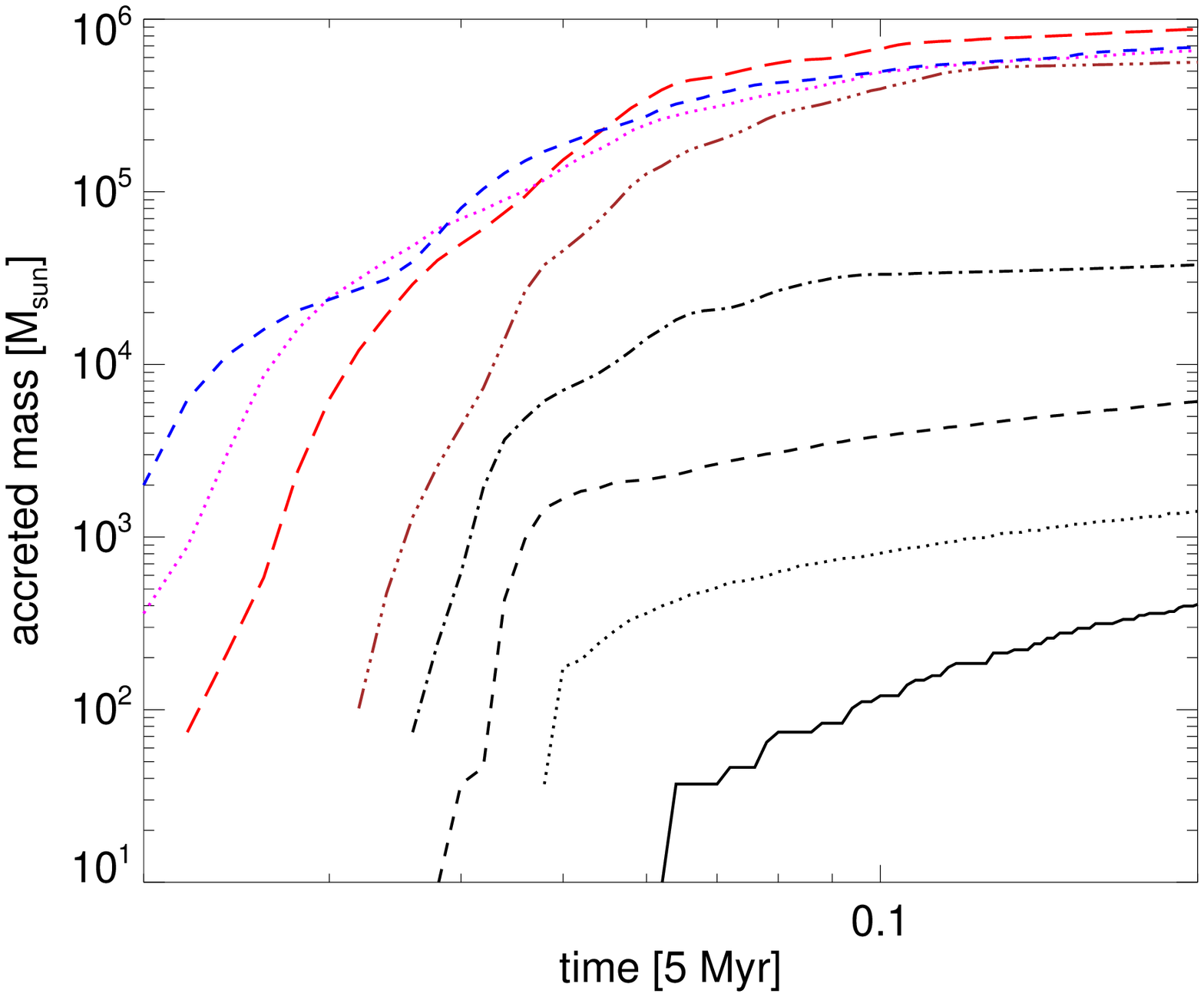,width=0.99\textwidth,angle=0}}
\end{minipage}
\begin{minipage}[h]{.48\textwidth}
\centerline{\psfig{file=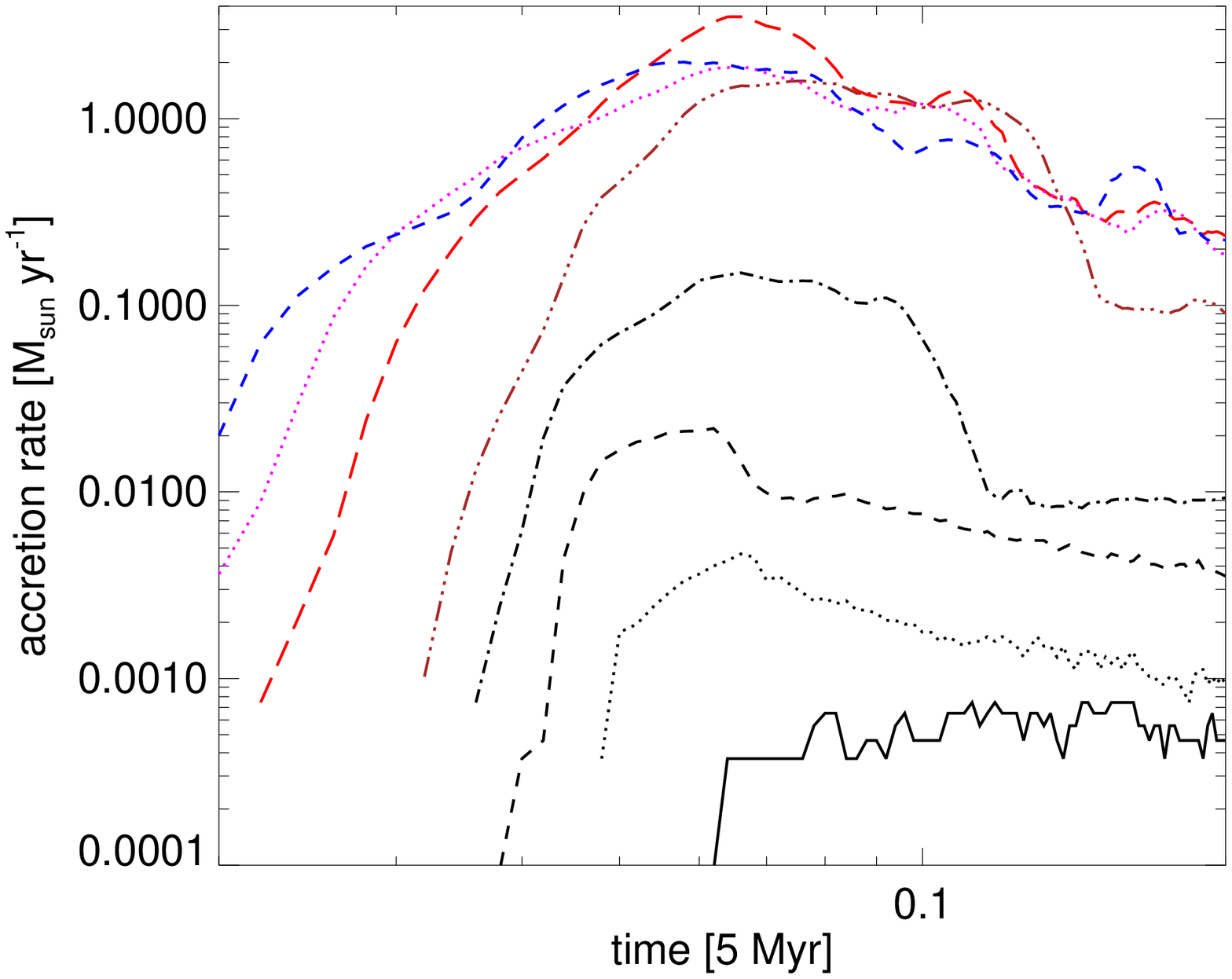,width=0.99\textwidth,angle=0}}
\end{minipage}
\caption{Accreted mass vs. time (left) and accretion rate vs. time (right) for
  a velocity field that contains high net rotation ($v_{\rm rot} =
  0.5$). Linestyles are as per Figure \ref{fig:vrot0}.}
\label{fig:vrot0.5}
\end{figure*}

\subsubsection{The highest rotation velocity case}

In the other extreme, Figure \ref{fig:vrot0.5} shows the accreted mass (left
panel) and the accretion rate (right panel) for the simulations S50 -- S57,
where the rotation velocity is set to the maximum explored in the paper, e.g.,
$v_{\rm rot} = 0.5$. The range of the turbulent velocity parameter and the
respective curve coding are the same as in Figure \ref{fig:vrot0}.

Comparing Figures \ref{fig:vrot0.5} and \ref{fig:vrot0}, it is obvious that
rotation significantly decreases the amount of gas accreted, and the accretion
rate. The reduction is most severe in the case of $v_{\rm turb} = 0$, by about
5 orders of magnitude. The role of turbulence in simulations S50 -- S57, as
suggested in \S \ref{sec:main_result}, is to increase the accretion rate. This
occurs by enhancement of the angular momentum distribution into the low
angular momentum end, and also by weakening the efficiency of shock mixing of
the gas.

\subsubsection{Rotation and turbulence parameter space}

We now combine the results of all our numerical experiments from S00 to S57 in
Table 1 by considering a single characteristic -- the accreted gas mass at
time $t = 0.2$ in code units. All of these runs have the same initial geometrical 
arrangement of gas and a fixed accretion radius $r_{\rm acc} = 10^{-3}$ but differ in the 
strengths of the initial turbulent and rotation velocities.

\begin{figure*}
\begin{minipage}[b]{.48\textwidth}
\centerline{\psfig{file=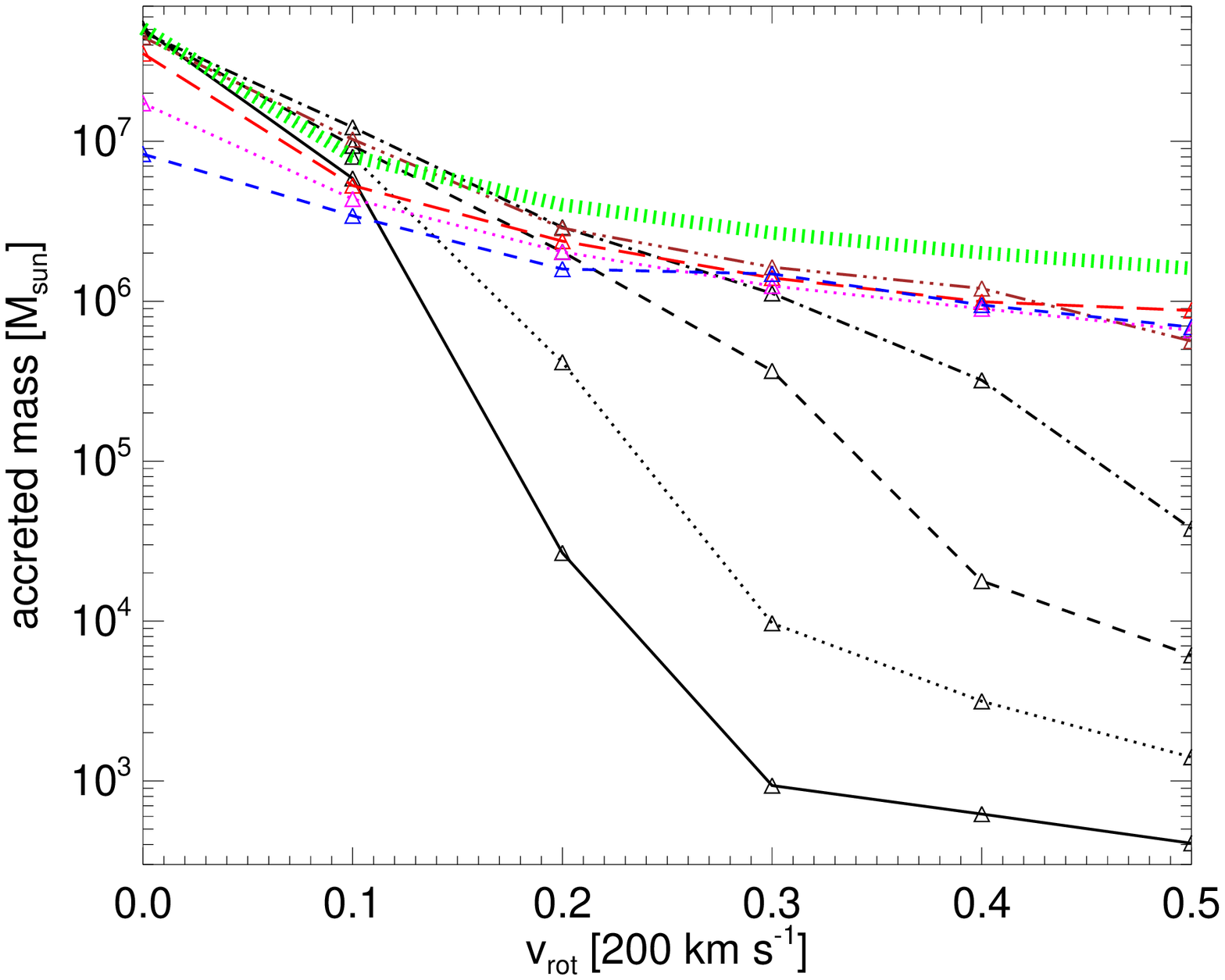,width=0.99\textwidth,angle=0}}
\end{minipage}
\begin{minipage}[b]{.48\textwidth}
\centerline{\psfig{file=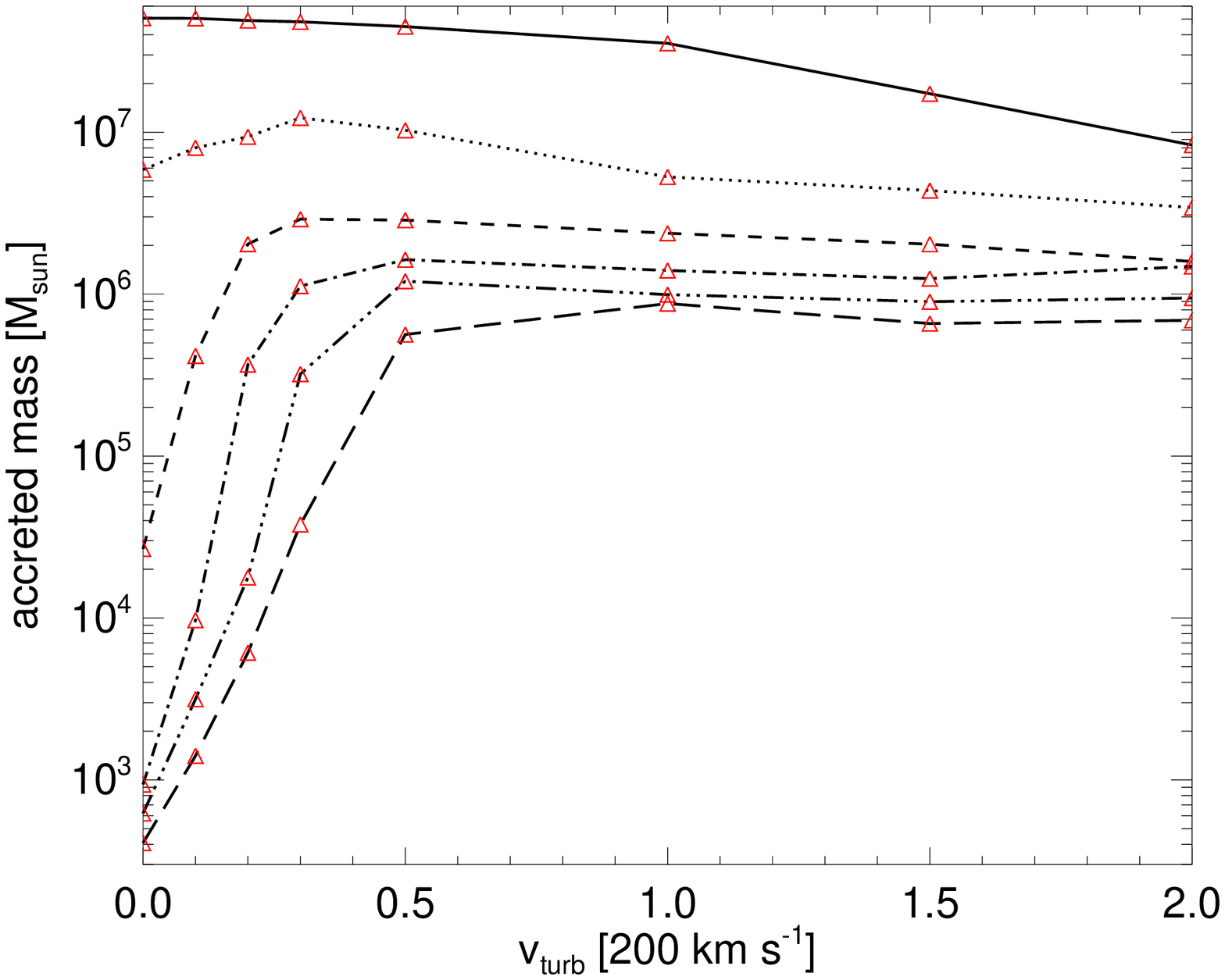,width=0.99\textwidth,angle=0}}
\end{minipage}
\caption{The accreted mass trend with rotation velocity, plotted
  for different levels of turbulence at $t=0.2$ (left) and the accreted mass trend
  with mean turbulent velocity, plotted for different levels of rotation at $t=0.2$
  (right).}
\label{fig:combinedtv}
\end{figure*}

The results are displayed in two ways. Figure \ref{fig:combinedtv} shows the accreted mass versus rotation velocity,
$v_{\rm rot}$. The simulations with the same level of mean turbulent velocity, $v_{\rm turb}$, are connected by the different lines. In particular,
the solid black curve shows the results for $v_{\rm turb} =0$, and the dashed
blue curve shows the results for $v_{\rm turb} = 2$. 

It is clear that increased rotation velocity, and thus net angular momentum,
always reduces the accreted mass, for any value of the turbulence parameter. 
This is in a general agreement with analytical expectations explored in \S
\ref{sec:vt_non_zero}. 

Together with the simulation results, we plot the predicted maximum accreted
mass (equation \ref{eq:mlc_equal}), which shows excellent agreement with the
maximum delineated by the levels of turbulence at saturation. The analytical
fit here is described specifically by
\begin{equation}
\frac{\Delta M}{M_{\rm shell}} = \textrm{min}\left[\frac{(GM_{\rm bh} r_{\rm
    acc})^{1/2}}{2 r_{\rm out}
  v_{\rm rot}}, 1\right]
\label{eq:mlc_equal_1}
\end{equation}
since the maximum fraction that can be accreted is unity and for $v_{\rm rot}
= 0$ we would expect the entire shell to have accreted by late times.

The increased level of turbulence acts to decrease the accretion rate 
in the case of very small rotation velocity $v_{\rm rot} \approx 0$ but
increase the accretion rate at higher rotation velocities (albeit up to saturation). This is not contradictory at 
all, however - the effect of
turbulence in all cases is to spread the initial angular momentum distribution
and prevent it from rapidly mixing into a single peak. In the case of no
initial rotation this reduces the accretion rate by moving gas from zero to
finite angular momentum orbits, whereas in the case of ``large'' initial net
angular momentum the accretion rate is increased as some gas is moved to the
low angular momentum orbits.

\subsubsection{Accretion and the size of the shell}\label{sec:outerradius}

Equation \ref{eq:mlc_shell} predicts that the fraction of accreted
mass should be inversely proportional to the shell's size, which we
characterise by $r_{\rm out}$. To test this prediction we repeated the
simulation S35 for a scaled-up version, $r_{\rm out} = 0.2$, $r_{\rm in} = 0.06$ and a scaled-down, $r_{\rm
  out} = 0.05$, $r_{\rm in} = 0.015$ shell (see runs S60 and S61 in Table 1). 

\begin{figure*}
\begin{minipage}[t]{.48\textwidth}
\centerline{\psfig{file=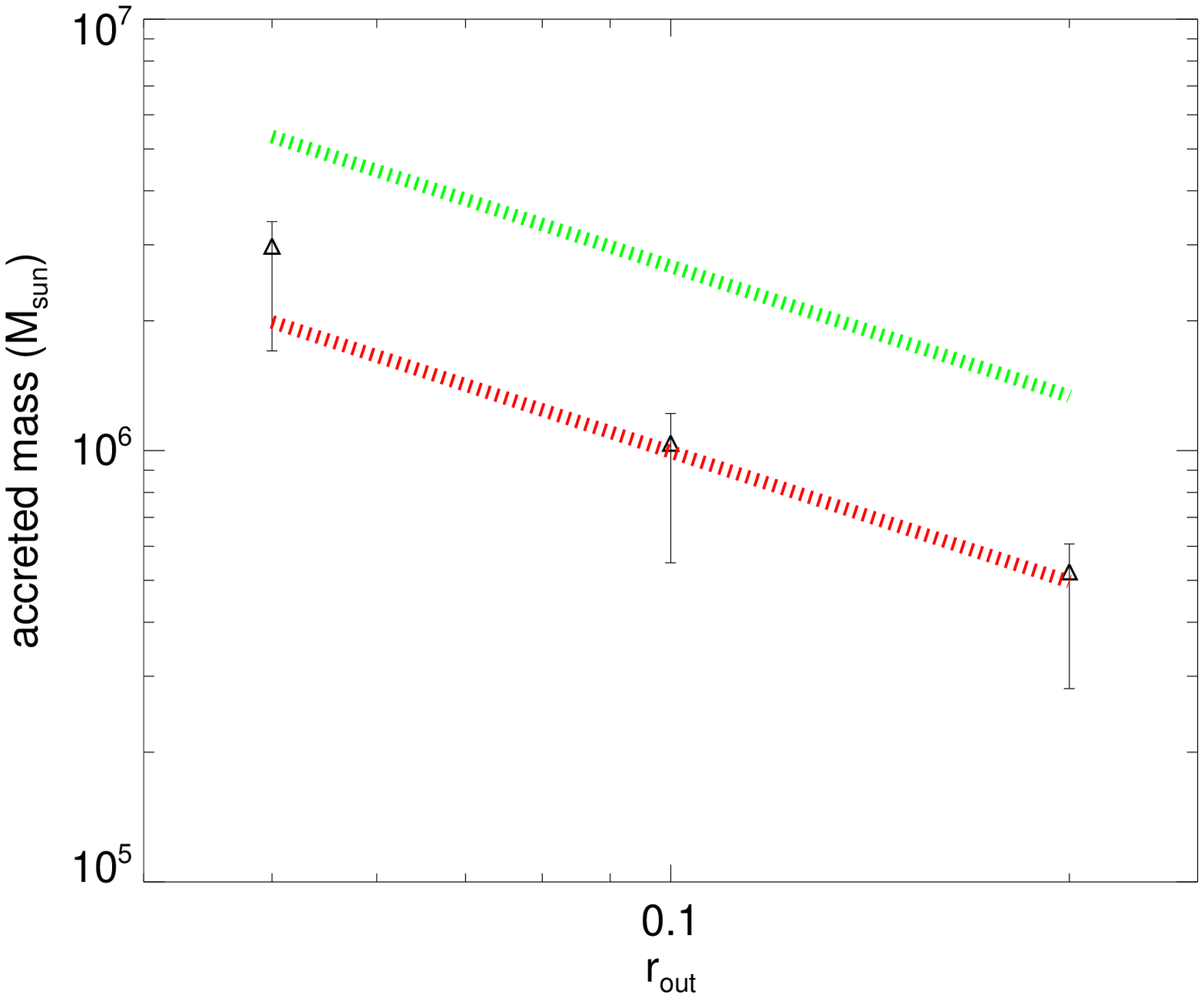,width=0.99\textwidth,angle=0}}
\end{minipage}
\begin{minipage}[b]{.48\textwidth}
\centerline{\psfig{file=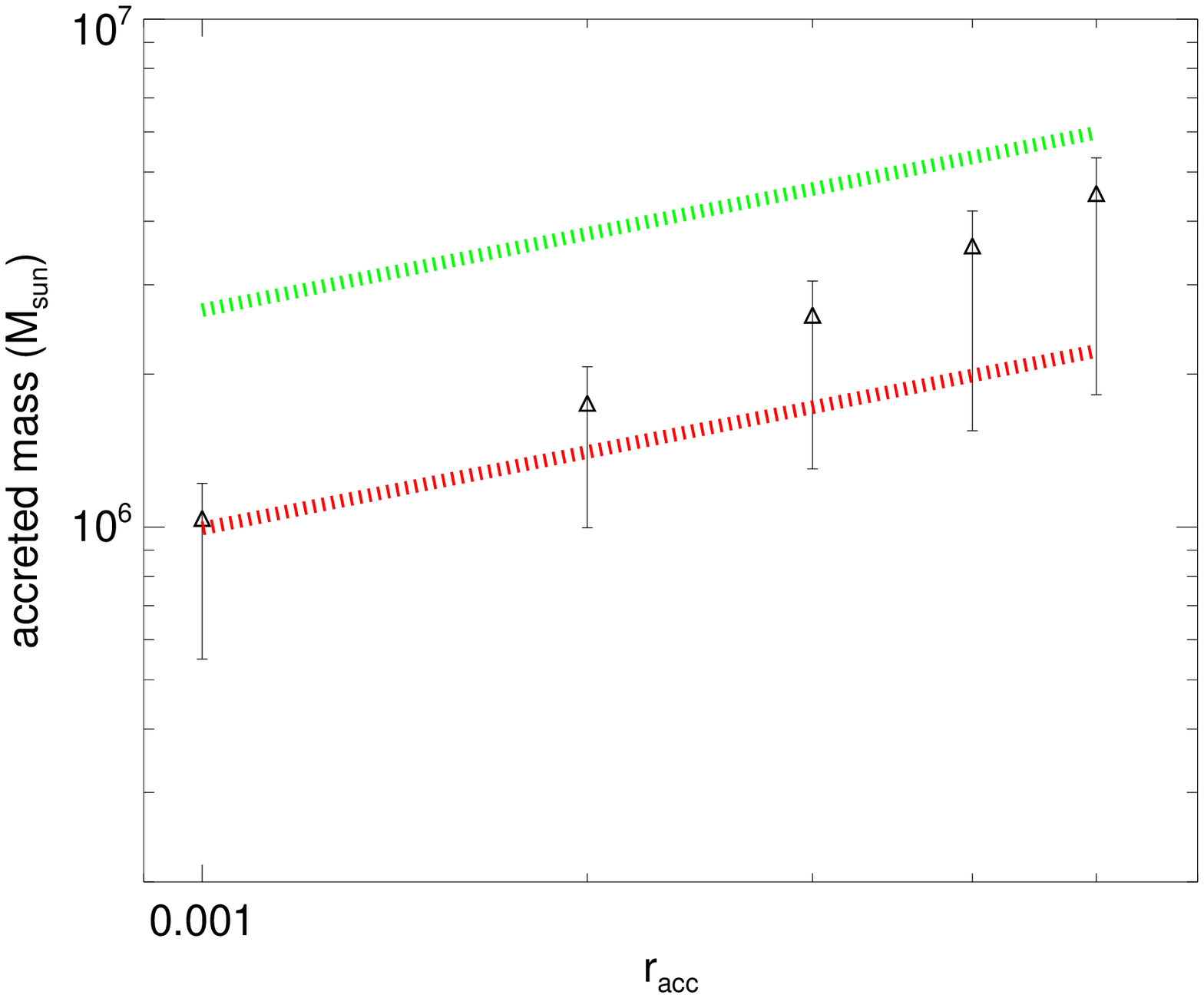,width=0.99\textwidth,angle=0}}
\end{minipage}
\caption{Scaling of accreted mass with $r_{\rm out}$ (left) and $r_{\rm acc}$
(right) for simulations with $v_{\rm rot} = 0.3$ and $v_{\rm turb} = 1.0$. Symbols correspond to the dynamical time at the
outer radius of the shell i.e., $t = 0.05$, $t = 0.1$, $t = 0.2$ (left) and
$t=0.1$ (right). Error bars are somewhat arbitrary, as it is not clear when
the main ballistic mode of accretion ends - here they denote $\pm
t/3$. Normalised analytical fits are equation \ref{eq:mlc_shell} (red), the
actual fit that would be expected for the simulation parameters, and equation
\ref{eq:mlc_equal} (green), the maximum accretion expected at saturation.}
\label{fig:outaccradius}
\end{figure*}

To be definitive, we compare the accreted gas mass at the dynamical time at the outer radius of the
shell, which corresponds to just after the initial peak in the accretion
rate. This is approximately the time at which we would expect the ballistic
mode to come to an end and a disc-dominated accretion mode to begin. The results are plotted
in Figure \ref{fig:outaccradius} (left panel) as a function of $r_{\rm out}$.

\subsubsection{Trend with accretion radius}

The accretion radius is a ``nuisance'' parameter of our study in the sense
that it is introduced only to allow the simulations to run in a reasonable
time. For physical reasons it would be desirable to make $r_{\rm acc}$ as
small as possible. Therefore it is important to test whether the suggested
analytical scaling of the results indeed holds.  With this in mind, we
repeated same simulation S35 for 5 different values of $r_{\rm acc}$, spanning
a range in $r_{\rm acc}$ from 0.001 to 0.005. The accreted mass from these simulations
is shown in Figure \ref{fig:outaccradius} (right panel). The red dotted power-law gives 
the analytical prediction. Although similar, it is somewhat less steep 
than the simulation results. 

\subsubsection{Visualisation of the circumnuclear disc}

\begin{figure*}
\begin{minipage}[b]{.48\textwidth}
\centerline{\psfig{file=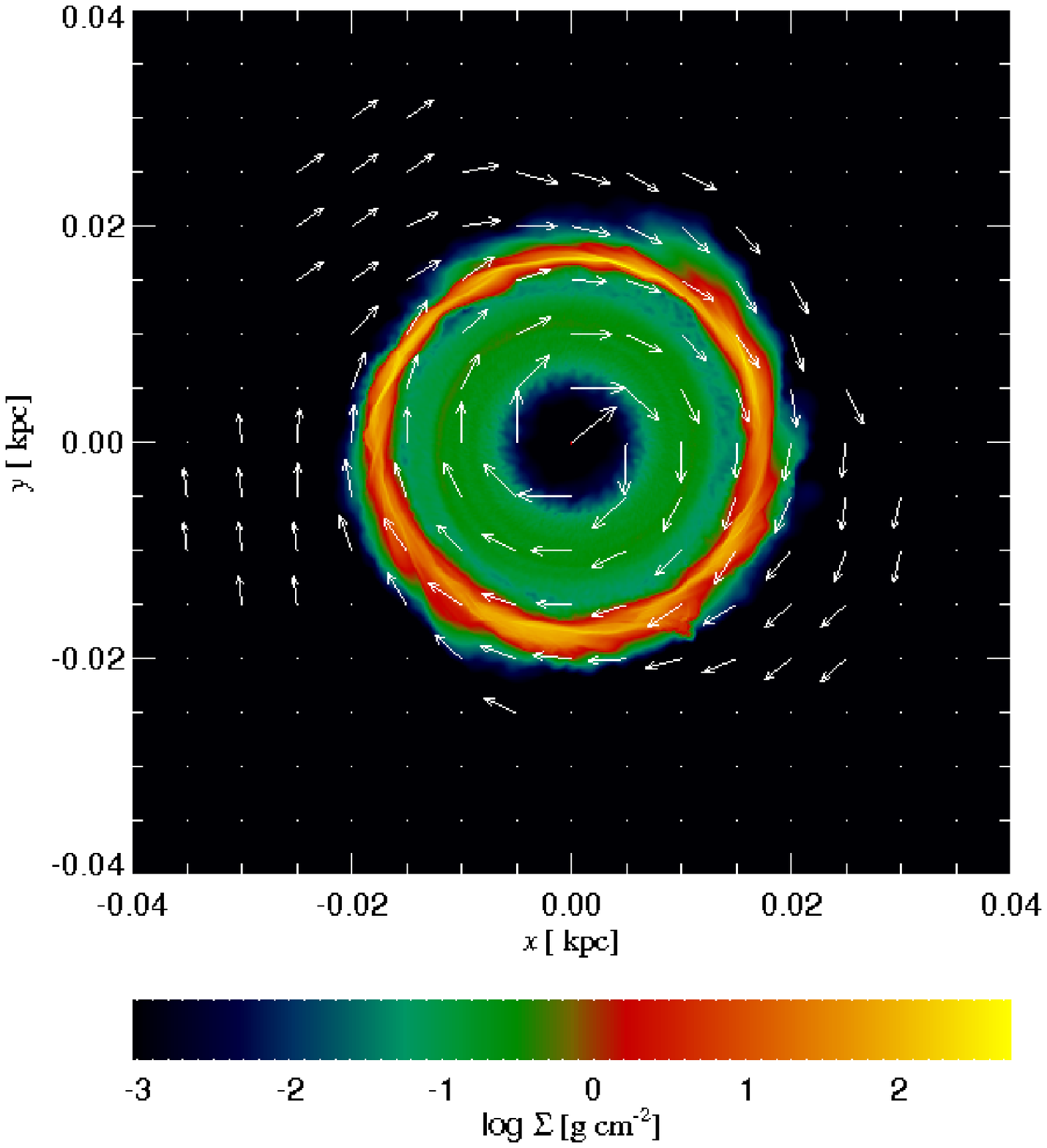,width=0.99\textwidth,angle=0}}
\end{minipage}
\begin{minipage}[b]{.48\textwidth}
\centerline{\psfig{file=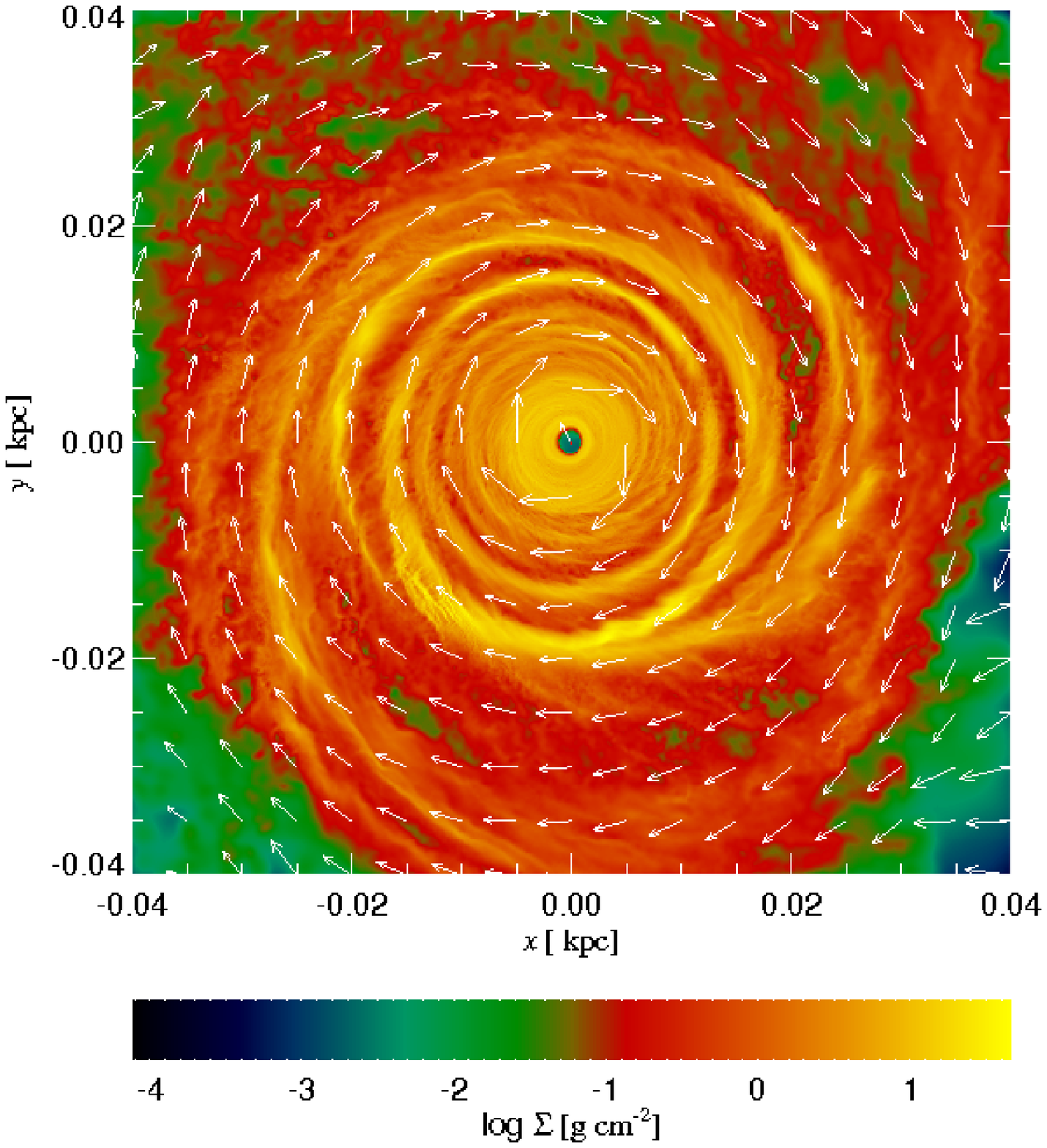,width=0.99\textwidth,angle=0}}
\end{minipage}
\caption{Face-on projection of the disc that forms by $t=0.418$ with no
  turbulence (left) and mean $v_{\rm turb} = 1.0$ (right). The initial
  rotation velocity for both was $v_{\rm rot} = 0.3$.}
\label{fig:CNDcomparison}
\begin{minipage}[b]{.48\textwidth}
\centerline{\psfig{file=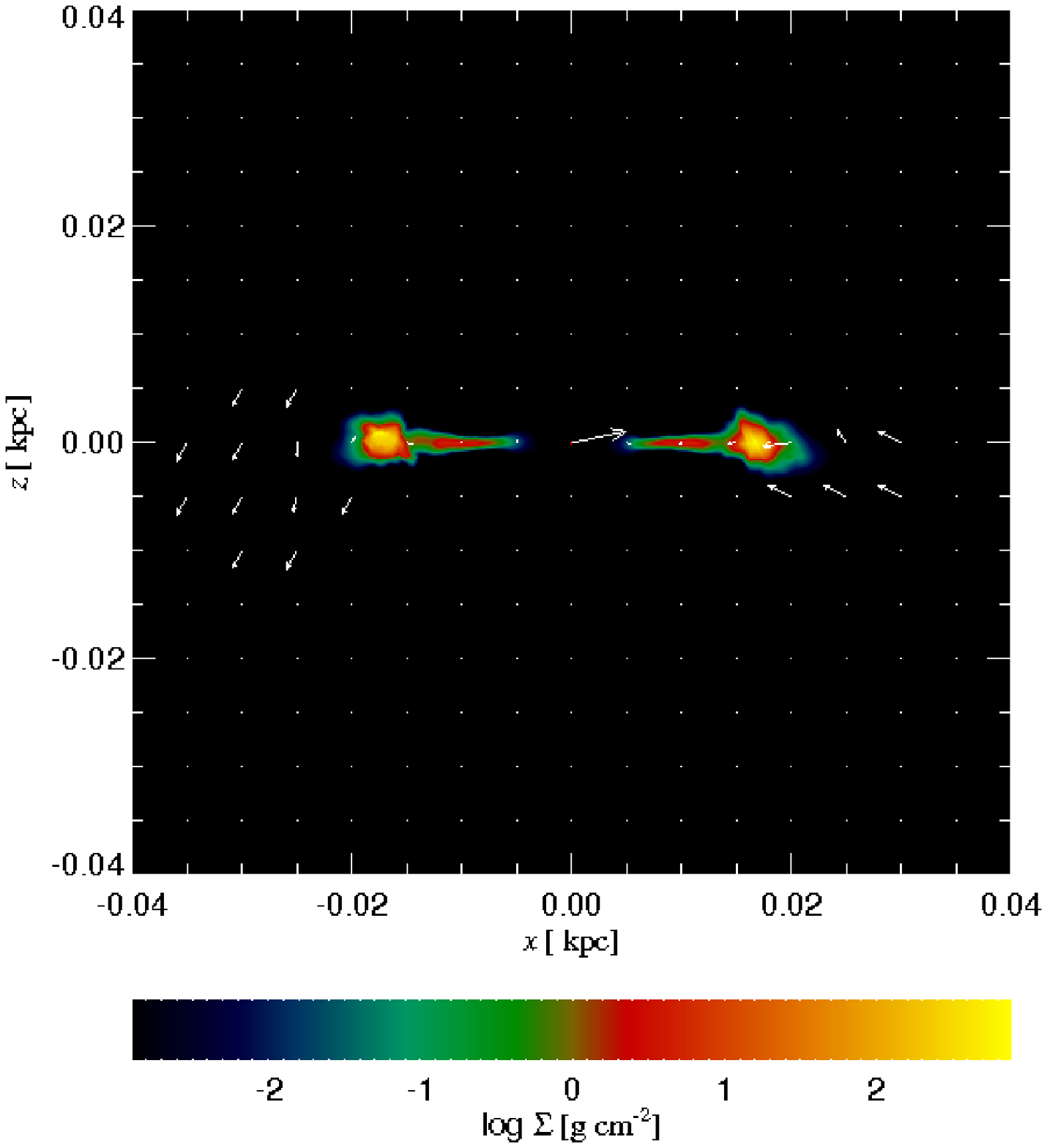,width=0.99\textwidth,angle=0}}
\end{minipage}
\begin{minipage}[b]{.48\textwidth}
\centerline{\psfig{file=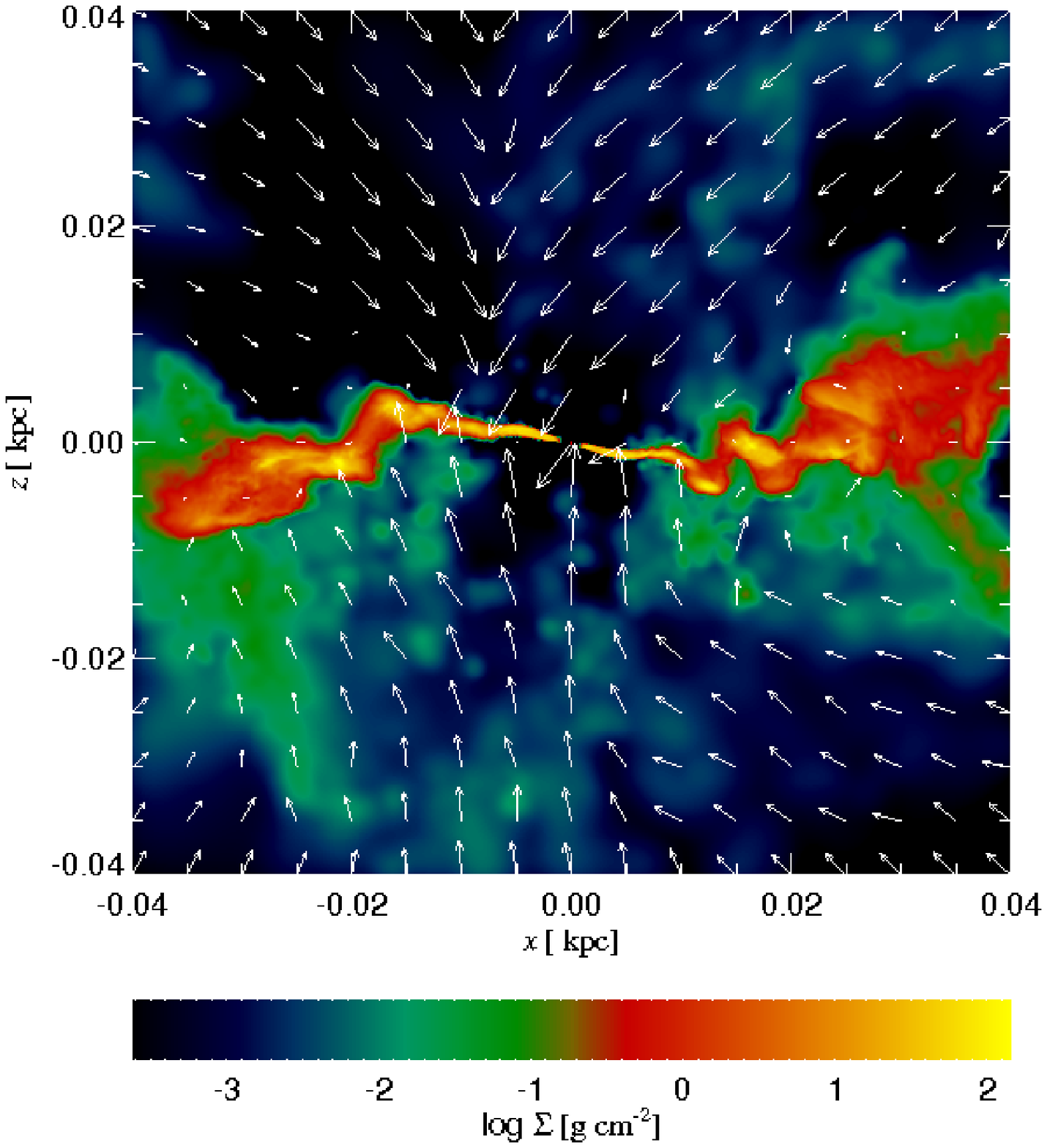,width=0.99\textwidth,angle=0}}
\end{minipage}
\caption{Slices of the CND (using $\tan \zeta = 2/5$) formed with no
  turbulence and with mean $v_{\rm turb} = 1.0$, both for
  $v_{\rm rot} = 0.3$ at t=0.418}
\label{fig:CNDcomparison90}
\end{figure*}

Figure \ref{fig:CNDcomparison} compares the surface density projected along
the $z$-axis for simulations S30 (left) and S35 (right) in the inner
$r = 0.04$ region.  Figure \ref{fig:CNDcomparison90} shows the same for these
two snapshots but projected along the $y$-axis.

As was already clear from Figure \ref{fig:lzgauss}, the gas forms a narrow
ring in the simulation S30. This occurs due to efficient mixing of
material with different angular momentum. This mixing is both radial (the ring
is narrower) and vertical (the ring is vertically less extended).

In contrast, in the simulation S35, turbulence creates a broad
distribution of angular momentum which does not mix so well. Mixing in the
vertical direction is actually reasonably efficient, as the disc is thin and
lies close to the $xy$ plane (see the right panel in Figure
\ref{fig:CNDcomparison90}). However the radial mixing is not so strong and the
disc radial structure is very extended in S35 compared with S30.

\subsubsection{Radial structure of the disc}\label{sec:radial}

Our analytical theory also makes a prediction for the surface density of the
disc in the innermost part. Indeed, the mass captured by the inner boundary of
radius $r_{\rm acc}$ scales as $M_{\rm capt} \propto r_{\rm acc}^{1/2}$. This
mass would likely sit in a rotationally supported disc of roughly $r\sim
r_{\rm acc}$ size. The surface density in the disc should then behave as
$\Sigma \propto M_{\rm capt}(r)/r^2 \propto r^{-3/2}$.

\begin{figure}
\begin{minipage}[b]{.48\textwidth}
\centerline{\psfig{file=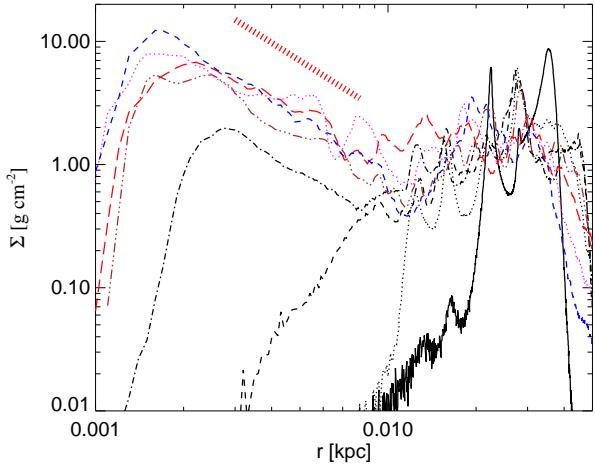,width=0.99\textwidth,angle=0}}
\end{minipage}
\caption{Azimuthally-averaged surface density versus radius for runs S50-S57
  at time $t=0.418$. Coloured curves correspond to simulations with different
  levels of mean turbulent velocity, as in Figure \ref{fig:vrot0}. The red
  line above the curves shows the theoretically expected (un-normalised) power law $\Sigma
  \propto r^{-3/2}$.}
\label{fig:sigmacombinedfit}
\end{figure}

This analytical prediction (red power-law) is compared to the simulation
results in Figure \ref{fig:sigmacombinedfit} which shows the surface density
defined on radial shells for all the different levels of turbulence for
$v_{\rm rot} = 0.5$. While the region beyound $r\sim 0.01$ shows separate
rings and highly variable surface density profiles for different simulations,
the disc inside the region $r\sim 0.01$ has a similar radial surface density
profile for all simulations with $v_{\rm turb} \simgt 0.5$. The agreement
between simulations and theory is quite reasonable in that region except for
runs with a small turbulent velocity $v_{\rm turb} < v_{\rm rot}$. For these
simulations the ballistic approximation is not appropriate and the gas does
not penetrate into the $r< 0.01$ region.  At radii larger than 0.01 the surface
density distribution is dominated by the interactions with the bulk of the
initial shell as it continues to fall in. The ballistic
approximation is of a limited use here because gas settling in the disc to
late times (i)
went through many collisions/interactions and hence the orbits are not
ballistic; and (ii) has a relatively large angular momentum, so the small
angle approximation we made in \S \ref{sec:circularisation} is violated.

The observed $\Sigma \propto r^{-3/2}$ dependence in the inner disc is another
piece of simple but convincing evidence that the ballistic accretion approximation that
we made in \S \ref{sec:vt_non_zero} is a reasonable one for the innermost
region of the flow.

\subsubsection{Vertical structure}\label{sec:vert}

Figure \ref{fig:CNDcomparison90} showed edge-on projections of the disc for
the simulation S30 (no turbulence) and (S35) initial turbulent velocity
$v_{\rm turb} = 1$. It is clear that the vertical stucture of the disc is
actually rather complex, especially beyond $r\sim 0.02$ region. This is
intuitively to be expected as the dynamical time is longest at larger radii, and
therefore circularisation and decay of vertical velocity dispersion in the
disc should take longer here. We can further quantify this by
assuming that the vertical velocity dispersion is pumped at the rate at which
the matter is brought in. Generation of momentum due to turbulence will therefore be
described by $\dot p_{\rm turb} \sim \dot\Sigma v_{\rm circ}$. These motions should decay on the local
dynamical time, $r/v_{\rm circ}$. Assuming a steady-state allows us to estimate the vertical velocity
dispersion as $\sigma_{\rm vert} \sim v_{\rm out} r/r_{\rm out}$, where
$v_{\rm out}$ is the circular velocity at $r_{\rm out}$. Therefore, the expected
disc aspect ratio in the outer parts (where the potential is isothermal) is $H/R \sim r/r_{\rm out} \propto
r$. 

\begin{figure}
\begin{minipage}[b]{.48\textwidth}
\centerline{\psfig{file=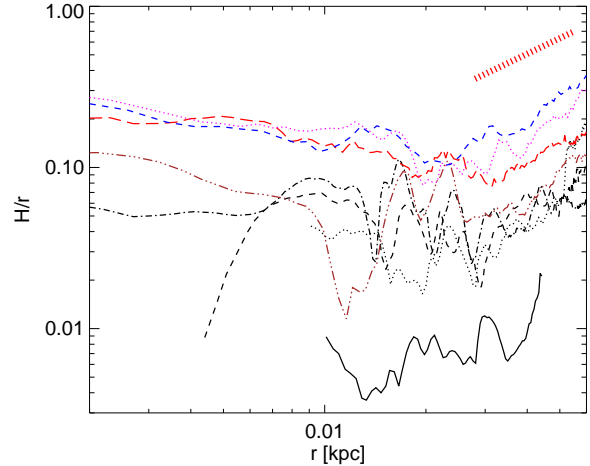,width=0.99\textwidth,angle=0}}
\end{minipage}
\caption{Dependence of vertical disc scaleheight, $H/r$, versus radius, for all
  levels of turbulence, with $v_{\rm rot} = 0.5$ (S50-S57), at time
  t=0.418. Line styles of the curves are the same as in Figure
  \ref{fig:vrot0}. The red line above the curves shows the theoretically
  expected result, for the (un-normalised) slope $H/r \propto r$.}
\label{fig:HoverR}
\end{figure}

We can calculate a dynamical, azimuthally-averaged disc scaleheight for the
circumnuclear disc that forms by using the projection of the
velocity dispersion along the `vertical' direction for each annulus i.e., the
mean angular momentum vector in each case. We therefore set,
\begin{equation}
H^2 = \frac{r^3 \sigma_{\rm vert}^2(r)}{GM_{\rm enc}(r)}
\label{eq:HoverR}
\end{equation}
where $M_{\rm enc}(r)$ is the enclosed mass at $r$, $\sigma_{\rm vert}(r)$ is the
`vertical' velocity dispersion at $r$. Figure \ref{fig:HoverR} plots ratios $H/r$ as functions of radius for simulations S50-S57. 
The solid line is for simulation S50, for which no initial turbulence is present. It is notable
that the disc formed by the flow in this case is geometrically much thinner than in all the cases
S51-S57. Among the turbulent cases, the general trend in $H/r$ is exactly as predicted above. 
The large scatter is explained by a strong time dependence: the disc is being stirred by strong 
waves that would eventually circularise the disc. 

In the region inside $r=0.01$, however, the observed dependence $H/r$ is flat
while the analytical expectation is $H/r \propto r^{3/2}$. We note that the
vertical disc scale height is much larger than the SPH smoothing length,
implying that the result inside $r=0.01$ is not due to under resolving of that
region.  We believe that the larger than expected vertical scale height is due
to dynamical heating of the inner disc by the waves already noted at larger
radii. Indeed, the inner disc is far less massive than the outer one, and any
waves present in the outer region have to either dissipate or get reflected in
this region. 

\section{Accretion inside the capture radius}


While we cannot resolve the hydrodynamics of gas within the accretion radius,
$r_{\rm acc}$ (1 pc for most of the simulations presented here), we track both
the mass and the angular momentum of the particles that are captured. We can
therefore calculate the mean angular momentum of the gas that settled inside
$r_{\rm acc}$. If the disc viscous time is comparable or longer than the
duration of our simulations, which is likely \citep[e.g.,][]{NC05}, then the
direction of the mean angular momentum of the gas captured within $r_{\rm
  acc}$ region is related to that of the sub-parsec disc. Note that in
principle the disc can be warped and then its structure is more complex
(Lodato et al 2009).

\begin{figure*}
\begin{minipage}[b]{.48\textwidth}
\centerline{\psfig{file=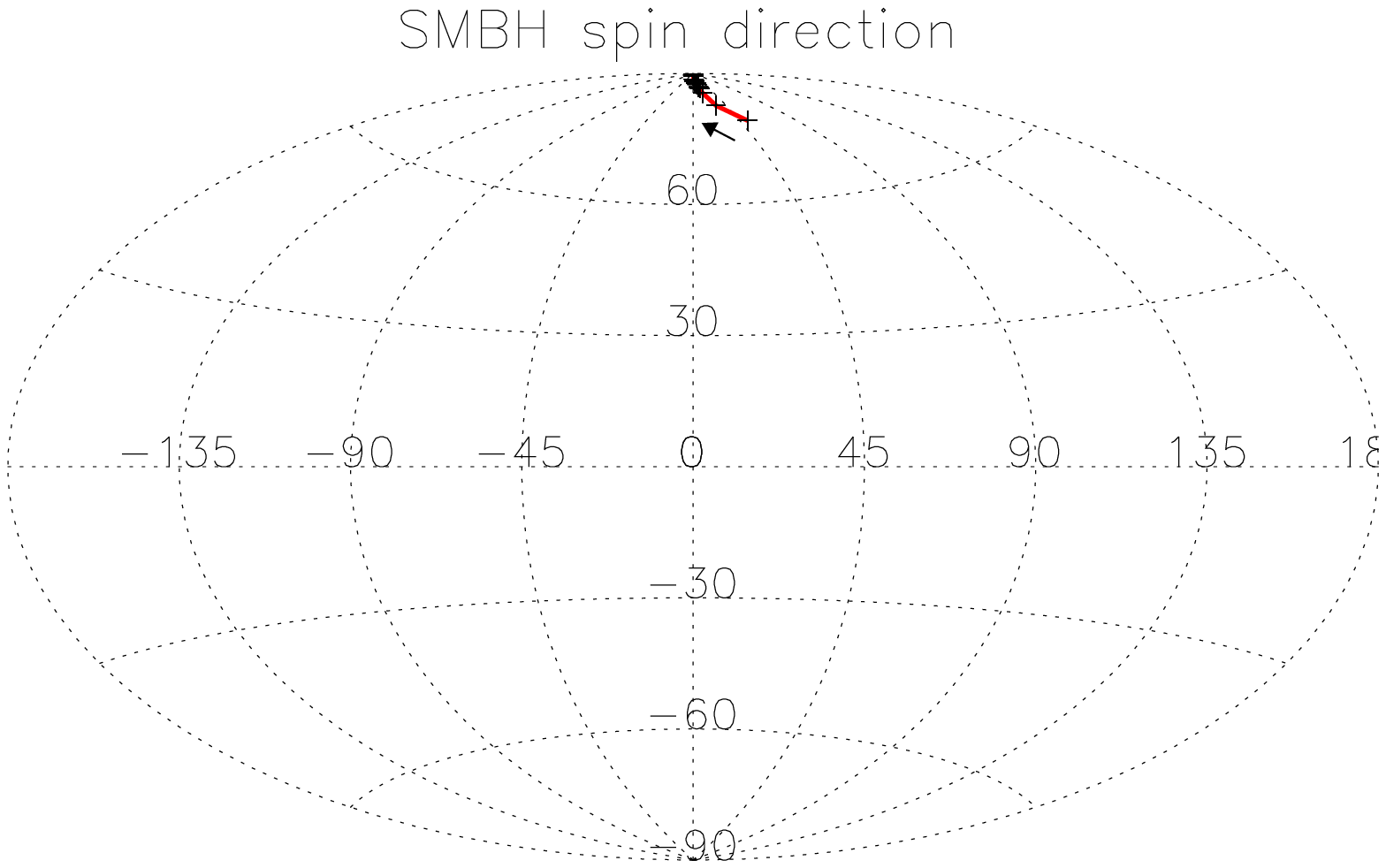,width=0.99\textwidth,angle=0}}
\end{minipage}
\begin{minipage}[b]{.48\textwidth}
\centerline{\psfig{file=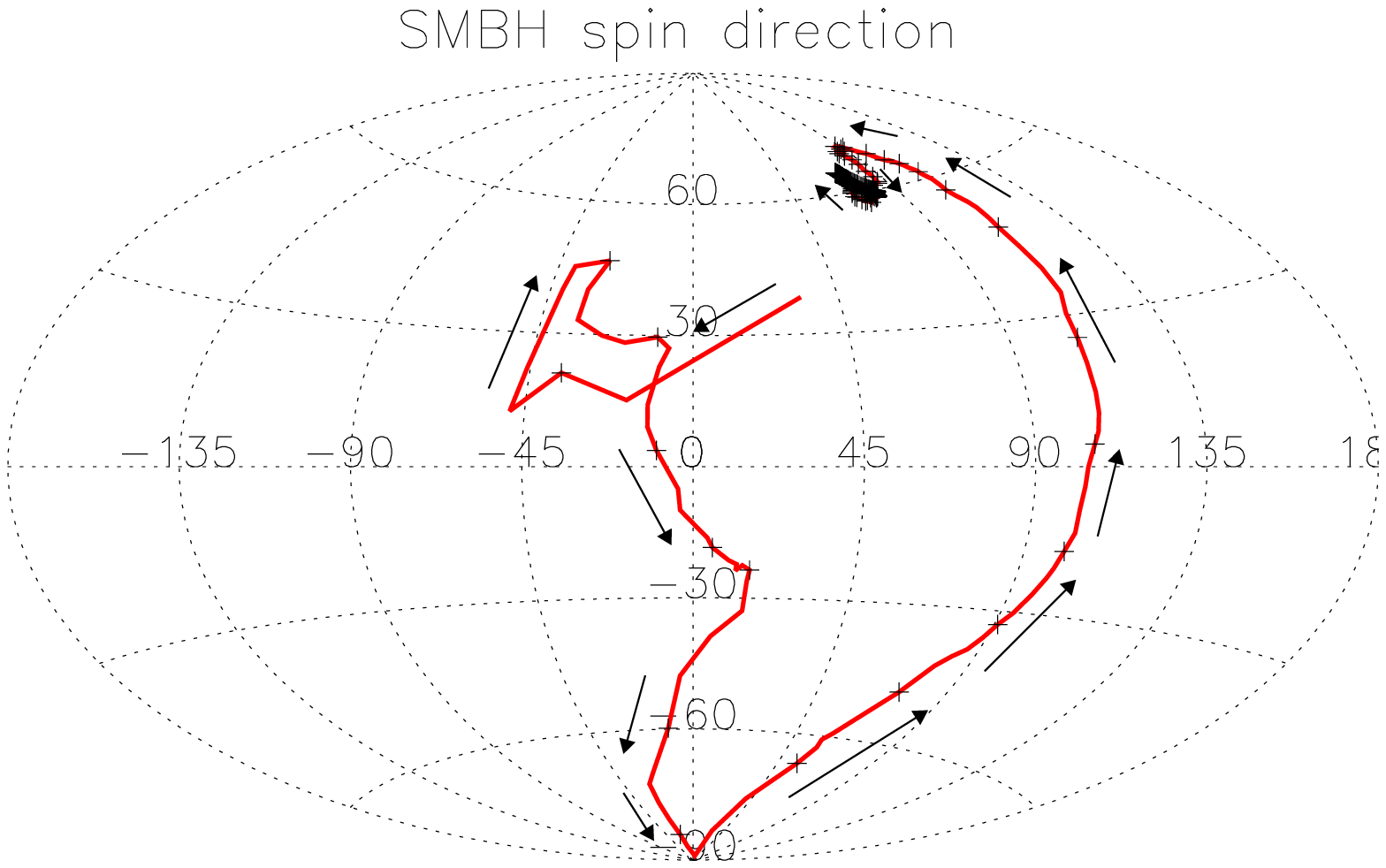,width=0.99\textwidth,angle=0}}
\end{minipage}
\caption{Plots of the accreted angular momentum direction in the
  Aitoff projection. Both plots are with $v_{\rm rot} = 0.3$, with no
  turbulence (left) and $v_{\rm turb} = 2.0$ (right). Crosses mark every 0.001
  in code time units. Note that the title of ``SMBH spin direction'' refers to the spin
  orientation that the SMBH would have if it had accreted all of the gas
  within the capture radius; in reality this is of course not the case.}
\label{fig:angular_momentum}
\end{figure*}

Figure \ref{fig:angular_momentum} shows the evolution of the direction of the
mean specific angular momentum for the gas particles accreted inside $r_{\rm
  acc}$ in the simulation S30 (left panel, no initial turbulence) and S37
(right panel, $v_{\rm turb} = 2$). The latter simulation is initialised with
the highest levels of turbulence we considered in this paper, and therefore
the effects of disc plane rotation are the strongest. The ``north pole'' in the diagram
corresponds to a specific angular momentum vector oriented along the
$z$-axis. Clearly, there is negligible deviation from that direction for the
``laminar'' simulation S30, as expected. For the simulation S37, however, there
is a complicated time evolution of the innermost disc direction. At one point
the disc appears to be completely counter-aligned to the mean angular momentum
of the shell.

\begin{figure}
\centerline{\psfig{file=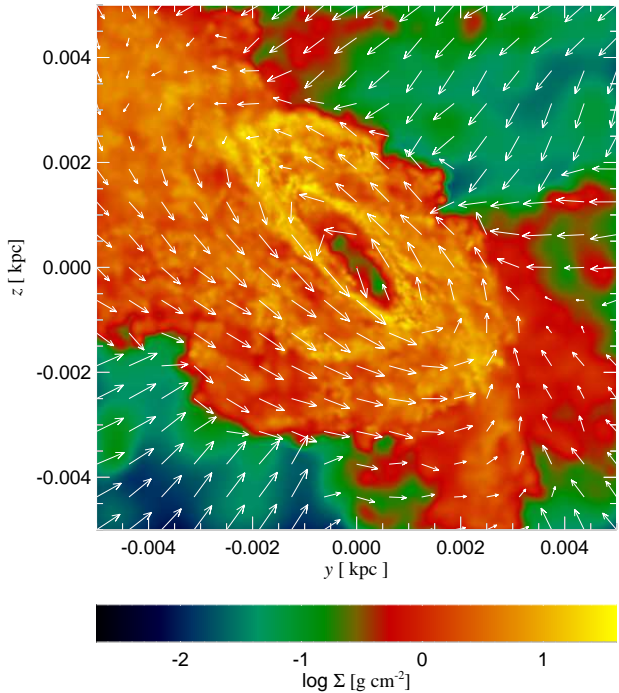,width=0.5\textwidth,angle=0}}
\caption{Projected gas density and velocity field for simulation S37, the
  innermost $r=0.005$ region, for  $t=0.072$. Note how strongly the disc is
  misaligned with the $xy$-plane. At later times the misalignment decreases
  as more material accumulates into the disc.}
\label{fig:vt2_vr0.3_fig2}
\end{figure}

The accretion disc ``plane wandering'' at $r \simlt r_{\rm acc}$ should also
be seen in the larger scale flow of gas.  Figure
\ref{fig:vt2_vr0.3_fig2} shows the innermost $r=0.005$ region for the
simulation S37 at time $t=0.072$. We can clearly see here that the disc in the immediate vicinity of $r_{\rm acc}$ is also
strongly displaced in terms of direction from the nominally expected one. One
should also note the flows of gas along directions roughly orthogonal to the disc
plane. These flows must have very different orientations of angular
momentum, and it is these that cause the plane of the
``sub-grid disc'' to execute the non-trivial path seen in Figure \ref{fig:angular_momentum}.

\section{Discussion}\label{sec:discussion}

\subsection{Summary of the results}

We have studied the hydrodynamics of an initially uniform rotating gaseous shell, seeded with an initial turbulent velocity spectrum as it falls into
the inner part of a spherically symmetric static potential. Particular
attention was paid to the amount of gas that made it inside the ``accretion
radius'' $r_{\rm acc}$.

In the control runs without initial turbulence, gas settles into a nearly
circular rotationally supported {\em ring} (see Figures \ref{fig:lzgauss},
\ref{fig:CNDcomparison}, \ref{fig:CNDcomparison90}). This was somewhat of a
surprise. Intuitively we expected a wide ($\Delta R \simgt R$) disc, as the
initial distribution of angular momentum is broad (Figure
\ref{fig:lzgauss}). Furthermore, only a few hundred solar masses of gas were
accreted, compared with the $\sim 5 \times 10^5 \msun$ expected on the basis
of a simple circularisation radius argument (\S \ref{sec:circularisation}). In
hindsight, both of these effects are caused by the effective mixing of gas by
the shocks. Low angular momentum gas gets mixed with high angular momentum gas
before the former reaches the capture radius $r=r_{\rm acc}$. The inner edge
of the ring is a significant distance from the accretion radius. In the context of AGN
accretion, such a ring would probably be strongly gravitationally unstable and
form stars rather than feed the SMBH
\citep[e.g.,][]{Goodman03,NayakshinEtal07}.

In contrast, runs in which the initial shell was seeded with turbulence show
qualitatively and quantitatively different results. The most significant of
these is that the accretion rates are higher by up to 3-4 orders of magnitude
compared with the simulations that had no initial turbulence imposed.  This is
particularly striking since the net angular momentum of these simulations is
the same.

It is obvious that turbulence broadens the angular momentum distribution,
setting some gas on low angular momentum orbits. A less trivial conclusion
from our numerical results is that turbulence also prevents the efficient gas
mixing pointed out above. Physically, high density regions are created by
supersonic turbulent convergent flows. These regions can then travel nearly
ballistically through the average density gas. Some of this gas is able to
reach the accretion radius, increasing the SMBH accretion rate compared with
runs without turbulence. This parameter space trend lasts until the initial mean turbulent velocity becomes greater than
the net rotation velocity, at which point the accretion trend saturates and
begins to slowly decrease as turbulence is increased further.

The development of high density regions in supersonic turbulence is by now
a classical result from a number of papers \citep[for reviews see
  e.g.,][]{ElmegreenScalo04,McKeeOstriker07}. There is also a body of work
that highlights the complicated dynamics of shocks in the multi-phase interstellar
medium. For example, \cite{Bonnell06,Dobbs08} suggest that shocks in a
multi-phase medium generate (or rather preserve) differential motions,
allowing the high density regions to slip through the shocks. These effects
are similar in nature to the ballistic behavior of high density clumps found
in our simulations.

\cite{SchartmannEtal09} simulated young nuclear star clusters injecting the
energy into the surrounding gas via supernovae and winds. As in our study it
was found that high density streams form. Those with small angular momentum
feed the small scale ``torus''. These results, and earlier results by
\cite{Cuadra06} for stellar winds in the central parsec of the Milky Way also support the viability of the ``ballistic'' AGN feeding mode.

We have attempted to build a simple analytical theory that would predict
the accretion rate for our capture inner boundary condition. We first
calculated the accretion rate in the ballistic approximation assuming that gas
in the shell receives monochromatic velocity kicks $v = v_{\rm turb}$ in
random directions in the frame rotating with velocity $v_{\rm rot}$. This
``monochromatic'' theory predicts that gas capture rate (fraction) is the
highest at $v_{\rm turb} \approx v_{\rm rot}$. We then took into account in an
approximate way the corrections associated with the fact that the real
turbulent velocity distribution function is not monochromatic. The fraction of
gas accreted within $r_{\rm acc}$ calculated in this way turned out to
describe our numerical results well. Furthermore, the theory predicted that
the inner disc surface density should have a power-law scaling: $\Sigma(r)
\propto r^{-3/2}$, which is indeed born out by the simulations. 

The approximate agreement of theory and simulations in this paper implies that
we can use our results for systems of astrophysical interest
by an appropriate rescaling of parameters, which we attempt to do in the next section.

\subsection{Can quasars and AGN be fed by ballistic
  accretion?}\label{sec:feeding}

The most serious challenge to theories of quasar and AGN feeding is
gravitational instability of these discs and the ensuing fragmentation of
discs into stars
\citep[e.g.,][]{Paczynski78,Kolykhalov80,CollinZahn99,Goodman03,NayakshinEtal07}.
Observations \citep[e.g.,][]{PaumardEtal06,BartkoEtal09} and numerical
simulations \citep{Bonnell08,HobbsNayakshin09} suggests that this process
operated in the central 0.5 pc of the Milky Way, creating young stars and
probably leaving the resident SMBH -- Sgr A* -- without gaseous fuel to shine
now \citep[e.g.,][]{Levin03,NC05}.

One obvious way in which the self-gravity ``catastrophe" of AGN discs can be
avoided is to deposit the gas at small enough radii
\citep[e.g.,][]{KingPringle07}. Namely, one can show that gaseous discs are
hot enough to avoid gravitational fragmentation within the self-gravity
radius, $R_{\rm sg}$, which is of the order of $0.01-0.1$ pc depending on the
SMBH mass and some of the disc parameters \citep[see][]{Goodman03}. Inclusion
of feedback from star formation allows the disc to survive and yet maintain an
interesting accretion rate on the SMBH out to a few times larger radii (Pedes
and Nayakshin, in preparation).

We shall now try to estimate the plausibility of such a small radius gas
deposition in the context of the ``ballistic'' gas deposition model we
have developed. To this end we shall use the approximate scalings for the fraction
of gas accreted from a rotating turbulent shell derived in this paper in \S
\ref{sec:vt_non_zero}. 

The size of the shell, $r_{\rm out}$, enters this estimate directly. Existence
of the $M_{\rm bh}-M_{\rm bulge}$ relation implies \citep[e.g.,][]{Haering04}
that the SMBH must be fed by gas located roughly the size of the bulge away
from the SMBH \citep{King03,King05}, or else it is difficult to see why the
quoted feedback link must exist \citep{NayakshinPower09}. We estimate that the
effective radius of the bulge, $R_b$, is about $2.5$ kpc for velocity
dispersion $\sigma = 150$ km/sec \cite[see][]{NayakshinEtal09}. If the rotation
velocity at $R_b$ is a small fraction of $\sigma$, say $v_{\rm rot} = 50$ km
s$^{-1}$, then the fraction of the gas that can be accreted from a turbulent shell is
\begin{equation}
\delta_{\rm max} \approx 0.001 GM_8^{1/2}
\left(\frac{r_{\rm
    acc}}{0.2\hbox{pc}}\right)^{1/2}\left(\frac{2.5\hbox{kpc}}{r_{\rm
    out}}\right) \left(\frac{50 \hbox{km s}^{-1}}{v_{\rm rot}}\right)
\label{eq:fractionlarge}
\end{equation}
where $M_8 = M_{\rm bh}/10^8\msun$. Here we have used equation
\ref{eq:mlc_equal}, assuming that the mean turbulent velocity is of the order
of the rotation velocity, which is our maximum in the accretion trend (although it should be noted that going to stronger turbulence
does not decrease the accreted mass by much - cf. \S \ref{sec:analysis}). The $M_{\rm bh} - M_{\rm bulge}$ relation
here gives the stellar mass in the bulge as $\sim 500 M_{\rm bh}$, and if we assume that $\sim$ a half of the
bulge gas was used up in star formation \citep{McGaughEtal09}, we find that the action of supernova
feedback in the context of our model yields an accreted mass of $\sim 5 \times
10^7 \msun$. This accretion would take place over roughly the dynamical time
at the outer edge of the bulge, giving us an accretion rate $\sim$ a few
$\msun $yr$^{-1}$. In this simple estimate, then, the SMBH
is indeed able to grow as massive as the observations require, assuming that
our ballistic accretion mode can be sustained for the required time. Future work should test
these ideas more directly, with ``turbulence'' driven directly by feedback
from star formation in the bulge, where the bulge gas distribution has been
derived from initial conditions in the gas on larger scales, e.g., the virial
radius of the host dark matter halo.

\subsection{A positive feedback link of star formation to AGN activity?}

Observations show that starburst and AGN activity are correlated in a number of ways
\citep[e.g.,][]{Fernandes01,GonzalesEtal01,FarrahEtal03}. One might think that
this is entirely natural, as both phenomena require a source of gas to be
present. However, star formation on kpc-scales does not have to be casually
connected with SMBH activity. As an example of this, consider a case when the angular
momentum of the gas is large, so that a kpc-scale disc is formed. If star
formation in this disc simply consumes the gas then there is nothing left to
feed the SMBH. This is what may have happened on smaller scales in the central parsec
of our Galaxy $\sim 6 $ million years ago
\citep{NC05,NayakshinEtal07}. Naturally, if such a process can rob the SMBH of
fuel at parsec scales then the situation becomes even worse at kiloparsec scales.

The opposite situation is also possible because the SMBH mass is a small
fraction of the bulge mass. It is not implausible to have a sub-pc scale, non
self-gravitating accretion disc that could feed a moderately bright AGN for a
long ($10^6-10^7$ years, say) time. The amount of gas involved in this could
be $10^{5}-10^7 \msun$, i.e., miniscule by galaxy standards. Why such activity
should always be connected with a powerful starburst is not clear.

However, the inclusion of star formation feedback could alter this
picture. For example, within a kpc-scale disc, feedback from stellar winds and supernovae
gives velocity kicks to the surrounding gas. For a standard IMF, the total momentum in 
supernova shells and winds from massive stars, integrated over their life
time, is about $\epsilon_* M_* c$, where $\epsilon \approx 10^{-3}$, and $M_*$ is
the total mass of the stellar population \citep{Leitherer1992,Thompson05}. If
this momentum is absorbed by gas with mass $M_g$ then the average kick
velocity for the gas is $v_{\rm kick} = (M_*/M_g) \epsilon c$, or $\sim 300$
km s$^{-1}$ for $M_* \sim M_g$. There is thus certainly enough momentum input
to make the gas highly turbulent. The cancellation of oppositely
directed momentum results in a reduction in this estimate. However, we have discussed here
only the momentum input; including supernova kinetic energy could act to increase the
above value \citep[e.g.,][]{DekelSilk86}. 

As we have shown in the paper, the effect of such feedback is likely to result in some small angular
momentum clouds or streams that may feed the SMBH. Our model therefore
predicts a positive correlation of starburst activity with SMBH activity, with
the latter being somewhat offset in time more often than not (it is also
possible to activate an AGN earlier if there is some low angular momentum gas in
the initial shell).

Further detailed physics-based modeling of AGN feeding, feedback, and star
formation, as well as comparisons to observations should shed light on the mode of
quasar feeding, and also on the mode of bulge formation.

\subsection{Cosmological cold stream SMBH feeding?}\label{sec:cold_stream}

We have employed turbulent initial conditions as a simple and mathematically convenient
way to introduce strong disordered differential motions in the gas for
controlled numerical experiments in AGN feeding.  We expect realistic galaxies
to be always in the ``turbulent'' regime in the above sense because the gas
contracting onto a dark matter halo is unlikely to be of uniform density or
possess a uniform velocity field. In fact, large-scale numerical simulations of galaxy formation
show cold gas streams that penetrate the shock-heated media of massive dark
matter haloes \citep{DekelEtal09}. This is physically similar to the ballistic
motion of gas we observed in our simulations. While it is very unlikely
that these very large-scale streams (hundreds of kpc) would be sufficiently well aimed
to actually feed the SMBH, collisions of several of such streams in the centre
of a galaxy would pump very strong differential motions and lead to some gas
being set on small angular momentum orbits. We therefore speculate that cold
large-scale gas streams may promote AGN feeding by providing energy
input to turbulence/differential motions on galactic scales.

\section{Conclusions}\label{sec:conclusion}

The classical challenge to AGN feeding is the large angular momentum of
gas expected to result in circularised discs of $\sim $ kpc scales (e.g.,
Shlosman, Frank \& Begelman 1989). Transporting gas from these scales is
extremely difficult due to long viscous times and the rapid consumption of gas
in star formation \citep[e.g.,][]{Goodman03,NayakshinEtal07}. Here we pointed
out that turbulence in the bulge, e.g., driven by supernova explosions, may be
a way to overcome the AGN feeding debacle by setting some gas on ballistic
orbits.

To test these ideas, we have performed simulations of hydrodynamical gas flow
with angular momentum, initially distributed in a spherical shell, in the
static spherical potential of a bulge with a central SMBH. The duration of our
simulations is a few dynamical times of the shell, allowing us to study the
formation of an accretion disc and the capture of gas within a small inner
boundary region. For all values of the rotation velocity, the
  shell was not in equilibrium with the background potential. This was done
  deliberately, as the goal of this project was to study the formation of a
  disc from an infalling gas distribution from first principles.

We found that angular momentum mixing in the runs without initial turbulence is very
effictive, resulting in an initial ring rather than an extended disc. Further
evolution of the ring depends on whether it can transfer angular momentum
quickly enough and avoid completely collapsing into stars. If star formation
consumes the ring quickly then such rings are a dead end as far as SMBH
feeding is concerned.

Turbulent motions in the shell overcome the angular momentum mixing problem by
creating high density regions that can travel nearly ballistically, retaining
their initial angular momentum. Provided there is enough gas with small
angular momentum the SMBH can be fed much more efficiently than in the
``laminar'' regime. The main conclusion here is that the star formation in the
bulge may not simply deplete the gas (hence depriving the SMBH of its fuel) but
may actually promote SMBH accretion by creating turbulent motions.

We note that turbulence in this particular project is just a mathematically convenient
formalism to introduce strong disordered differential motions in the gas. We
expect that realistic galaxies are always in the turbulent regime because the
state of the gas on large scales is unlikely to be that of uniform density and
angular momentum.

Further work with self-consistently driven turbulence from star formation and
realistic initial (large-scale) boundary conditions is needed to establish the
relevance of the ``ballistic" accretion to feeding the SMBH, although our
research suggests that it is a promising model.

\section{Acknowledgments}

Helpful discussions with Rashid Sunyaev, Eugene Churazov, Simon White, Walter
Dehnen, Justin Read and Peter Cossins are gratefully acknowledged. Theoretical Astrophysics
group at the University of Leicester is funded by a research grant from
STFC. AH acknowledges an STFC studentship.

\bibliographystyle{mnras} 

\bibliography{nayakshinalex_1701}

\appendix

\begin{table*}
\caption{Initial conditions for each simulation. The ratio with respect to the gravitational potential
energy gives an indication of how well supported the shell is against the
external potential through virial motions: $(E_{\rm turb} + E_{\rm therm})/|E_{\rm grav}| \simgt 1/2$ for a virialised shell.}
\centering
\begin{tabular}{|c|c|c|c|c|c|c|c|c|c|c|c|c|c|c|c|c|}\hline
ID & $r_{\rm in}$ & $r_{\rm out}$ & $v_{\rm rot}$ & $v_{\rm turb}$ & $\frac{E_{\rm turb} + E_{\rm therm}}{<|E_{\rm
    grav}|>}$ & $<r_{\rm circ}>^{*}$ & $r_{\rm acc}$ & $M_{\rm acc}
(\msun)^{1}$ \\ \hline
\hline
S00 & 0.03 & 0.1 & 0.0 & 0.0 &$6 \times 10^{-5}$ & 0.0 & 0.001 &$ 5.01 \times 10^{7}$ \\ 
S01 & 0.03 & 0.1 & 0.0 & 0.1 & 0.001 & 0.0 &0.001 &$ 5.01 \times 10^{7}$ \\ 
S02 & 0.03 & 0.1 & 0.0 & 0.2 & 0.004 & 0.0 & 0.001 &$ 4.90 \times 10^{7}$ \\ 
S03 & 0.03 & 0.1 & 0.0 & 0.3 & 0.009 & 0.0 & 0.001 &$ 4.81 \times 10^{7}$ \\ 
S04 & 0.03 & 0.1 & 0.0 & 0.5 & 0.026 & 0.0 & 0.001 &$ 4.49 \times 10^{7}$ \\ 
S05 & 0.03 & 0.1 & 0.0 & 1.0 & 0.103 & 0.0 & 0.001 &$ 3.53 \times 10^{7}$ \\ 
S06 & 0.03 & 0.1 & 0.0 & 1.5 & 0.231 & 0.0 & 0.001 &$ 1.72 \times 10^{7}$ \\ 
S07 & 0.03 & 0.1 & 0.0 & 2.0 & 0.410 & 0.0 & 0.001 &$ 8.40 \times 10^{6}$ \\ 
\\
S10 & 0.03 & 0.1 & 0.1 & 0.0 & $6 \times 10^{-5}$ & 0.004 & 0.001 & $ 5.86 \times 10^{6}$ \\ 
S11 & 0.03 & 0.1 & 0.1 & 0.1 & 0.001 & 0.004 & 0.001 & $ 8.01 \times 10^{6}$ \\ 
S12 & 0.03 & 0.1 & 0.1 & 0.2 & 0.004 & 0.004 & 0.001 & $ 9.38 \times 10^{6}$ \\ 
S13 & 0.03 & 0.1 & 0.1 & 0.3 & 0.009 & 0.004 & 0.001 & $ 1.23 \times 10^{7}$ \\ 
S14 & 0.03 & 0.1 & 0.1 & 0.5 & 0.026 & 0.004 & 0.001 & $ 1.03 \times 10^{7}$ \\ 
S15 & 0.03 & 0.1 & 0.1 & 1.0 & 0.103 & 0.004 & 0.001 & $ 5.29 \times 10^{6}$ \\ 
S16 & 0.03 & 0.1 & 0.1 & 1.5 & 0.231 & 0.004 & 0.001 & $ 4.36 \times 10^{6}$ \\ 
S17 & 0.03 & 0.1 & 0.1 & 2.0 & 0.410 & 0.004 & 0.001 & $ 3.40 \times 10^{6}$ \\ 
\\
S20 & 0.03 & 0.1 & 0.2 & 0.0 & $6 \times 10^{-5}$ & 0.011 & 0.001 & $ 2.67 \times 10^{4}$ \\  
S21 & 0.03 & 0.1 & 0.2 & 0.1 & 0.001 & 0.011 & 0.001 & $ 4.16 \times 10^{5}$ \\ 
S22 & 0.03 & 0.1 & 0.2 & 0.2 & 0.004 & 0.011 & 0.001 & $ 2.04 \times 10^{6}$ \\ 
S23 & 0.03 & 0.1 & 0.2 & 0.3 & 0.009 & 0.011 & 0.001 & $ 2.90 \times 10^{6}$ \\
S24 & 0.03 & 0.1 & 0.2 & 0.5 & 0.026 & 0.011 & 0.001 & $ 2.87 \times 10^{6}$ \\ 
S25 & 0.03 & 0.1 & 0.2 & 1.0 & 0.103 & 0.011 & 0.001 & $ 2.37 \times 10^{6}$ \\ 
S26 & 0.03 & 0.1 & 0.2 & 1.5 & 0.231 & 0.011 & 0.001 & $ 2.04 \times 10^{6}$ \\ 
S27 & 0.03 & 0.1 & 0.2 & 2.0 & 0.410 & 0.011 & 0.001 & $ 1.58 \times 10^{6}$ \\ 
\\
S30 & 0.03 & 0.1 & 0.3 & 0.0 & $6 \times 10^{-5}$ & 0.019 & 0.001 & $ 9.35
\times 10^{2}$ \\  
S31 & 0.03 & 0.1 & 0.3 & 0.1 & 0.001 & 0.019 & 0.001 & $ 9.69 \times 10^{3}$ \\ 
S32 & 0.03 & 0.1 & 0.3 & 0.2 & 0.004 & 0.019 & 0.001 & $ 3.67 \times 10^{5}$ \\ 
S33 & 0.03 & 0.1 & 0.3 & 0.3 & 0.009 & 0.019 & 0.001 & $ 1.12 \times 10^{6}$ \\ 
S34 & 0.03 & 0.1 & 0.3 & 0.5 & 0.026 & 0.019 & 0.001 & $ 1.63 \times 10^{6}$ \\ 
S35 & 0.03 & 0.1 & 0.3 & 1.0 & 0.103 & 0.019 & 0.001 & $ 1.40 \times 10^{6}$ \\ 
S36 & 0.03 & 0.1 & 0.3 & 1.5 & 0.231 & 0.019 & 0.001 & $ 1.25 \times 10^{6}$ \\ 
S37 & 0.03 & 0.1 & 0.3 & 2.0 & 0.410 & 0.019 & 0.001 & $ 1.48 \times 10^{6}$ \\ 
\\
S40 & 0.03 & 0.1 & 0.4 & 0.0 & $6 \times 10^{-5}$ & 0.026 & 0.001 & $ 6.21
\times 10^{2} $ \\  
S41 & 0.03 & 0.1 & 0.4 & 0.1 & 0.001 & 0.026 & 0.001 & $ 3.15 \times 10^{3}$ \\ 
S42 & 0.03 & 0.1 & 0.4 & 0.2 & 0.004 & 0.026 & 0.001 & $ 1.78 \times 10^{4}$ \\ 
S43 & 0.03 & 0.1 & 0.4 & 0.3 & 0.009 & 0.026 & 0.001 & $ 3.21 \times 10^{5}$ \\ 
S44 & 0.03 & 0.1 & 0.4 & 0.5 & 0.026 & 0.026 & 0.001 & $ 1.20 \times 10^{6}$ \\ 
S45 & 0.03 & 0.1 & 0.4 & 1.0 & 0.103 & 0.026 & 0.001 & $ 9.93 \times 10^{5}$ \\ 
S46 & 0.03 & 0.1 & 0.4 & 1.5 & 0.231 & 0.026 & 0.001 & $ 8.98 \times 10^{5}$ \\ 
S47 & 0.03 & 0.1 & 0.4 & 2.0 & 0.410 & 0.026 & 0.001 & $ 9.47 \times 10^{5}$ \\ 
\\
S50 & 0.03 & 0.1 & 0.5 & 0.0 & $6 \times 10^{-5}$ & 0.034 & 0.001 & $ 4.08
\times 10^{2}$ \\  
S51 & 0.03 & 0.1 & 0.5 & 0.1 & 0.001 & 0.034 & 0.001 & $ 1.41 \times 10^{3}$ \\ 
S52 & 0.03 & 0.1 & 0.5 & 0.2 & 0.004 & 0.034 & 0.001 & $ 6.10 \times 10^{3}$ \\ 
S53 & 0.03 & 0.1 & 0.5 & 0.3 & 0.009 & 0.034 & 0.001 & $ 3.78 \times 10^{4}$ \\ 
S54 & 0.03 & 0.1 & 0.5 & 0.5 & 0.026 & 0.034 & 0.001 & $ 5.63 \times 10^{5}$ \\ 
S55 & 0.03 & 0.1 & 0.5 & 1.0 & 0.103 & 0.034 & 0.001 & $ 8.75 \times 10^{5}$ \\ 
S56 & 0.03 & 0.1 & 0.5 & 1.5 & 0.231 & 0.034 & 0.001 & $ 6.58 \times 10^{5}$ \\ 
S57 & 0.03 & 0.1 & 0.5 & 2.0 & 0.410 & 0.034 & 0.001 & $ 6.89 \times 10^{5}$ \\ 
\\
S60 & 0.015 & 0.05 & 0.3 & 1.0 & 0.091 & 0.008 & 0.001 &$ 4.43 \times 10^{6}$ \\
S61 & 0.06 & 0.2 & 0.3 & 1.0 & 0.119 & 0.041 & 0.001 &$ 5.23 \times 10^{5}$ \\
\\
S70 & 0.03 & 0.1 & 0.3 & 1.0 & 0.103 & 0.019 & 0.002 & $ 2.36 \times 10^{6}$ \\
S71 & 0.03 & 0.1 & 0.3 & 1.0 & 0.103 & 0.019 & 0.003 & $ 3.45 \times 10^{6}$ \\
S72 & 0.03 & 0.1 & 0.3 & 1.0 & 0.103 & 0.019 & 0.004 & $ 4.77 \times 10^{6}$ \\
S73 & 0.03 & 0.1 & 0.3 & 1.0 & 0.103 & 0.019 & 0.005 & $ 6.07 \times 10^{6}$ \\

\hline
\end{tabular}
\begin{flushleft}
$^{*}$$r_{\rm circ}$ here refers to the initial value of circularisation
radius for each of the particles, calculated in the region of the potential
that they start in, based solely on the rotation velocity.

$^{1}$ value of mass accreted taken at $t=0.2$ in code units.
\end{flushleft}
\label{table1}
\end{table*}

\section{An alternative velocity distribution for the shell}

On the referee's request, we have re-simulated a fiducial parameter set with
an alternative velocity field. In this case we set $l$, the specific angular momentum, to be
constant, rather than the rotation velocity. The magnitude of the velocity at a given
radius is then
\begin{equation}
v_\phi = \frac{l_{\rm rot}}{r}
\end{equation}
where $l_{\rm rot}$ is a constant. This is a more conservative initial
condition in the sense that this distribution would be \emph{expected} to form an
infinitely thin ring, since the distribution is a delta function in angular
momentum space. The broadening of this distribution due to turbulence would
therefore be more difficult, and any enhancement to the accretion rate that
this broadening provides is therefore a significant result. We note that the other ``natural'' choice of velocity field
for a spherical shell, namely solid body rotation, is only
reasonable for a system in hydrostatic equilibrium, which our shell is not.

\begin{figure}
\centerline{\psfig{file=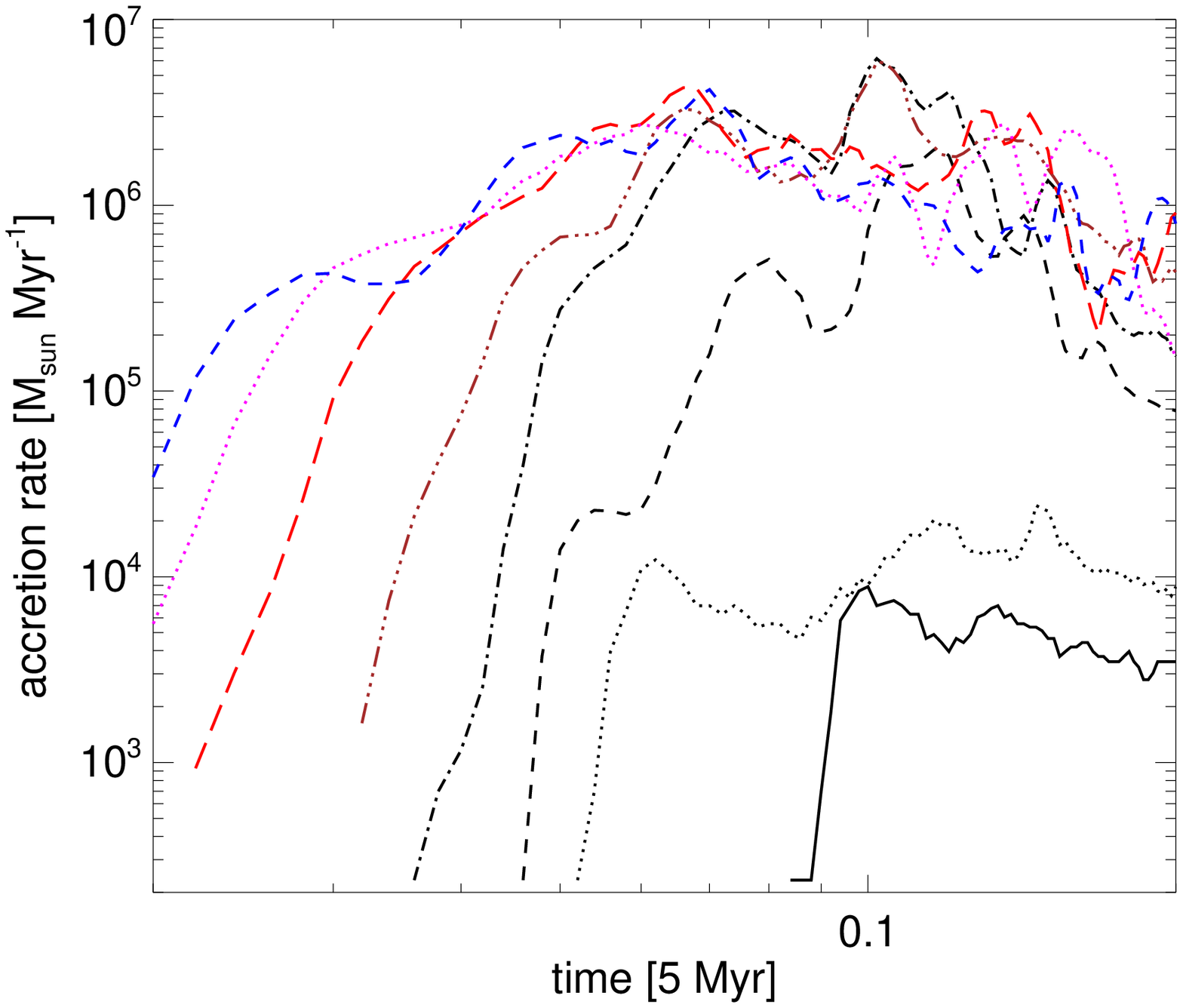,width=0.5\textwidth,angle=0}}
\caption{Accretion rate versus time for a set of runs with a constant angular
  momentum velocity field, with $l_{\rm rot} = 0.02$ in code units. The other parameters
  for these runs were held at the fiducial values of S30, and color/linestyles
  is as standard for the other plots in the paper. The main result of
  the paper still holds: the accretion rate on the
  black hole strongly increases with increasing levels of turbulence when
  rotation is present (see \S \ref{sec:main_result}).}
\label{fig:lconst}
\end{figure}

As can be seen from Figure \ref{fig:lconst}, our main conclusion is unchanged
for this alternative velocity distribution. The accretion rate onto the inner
parsec is increased by $\sim 2$ orders of magnitude from zero/low turbulence to
high turbulence. We note the the constant angular momentum within a spherical
shell is, too, a strongly idealised setup, with similar behaviour at the polar
regions to that of the constant $v_{\rm rot}$ condition. However, we are again
attempting to model a simple case of gas mixing due to different angular
momentum at similar radii, and unfortunately there is no ``realistic''
single-parameter velocity field that can describe a spherical distribution
with these properties.

\section{Convergence tests on numerical viscosity and resolution}

\begin{figure}
\centerline{\psfig{file=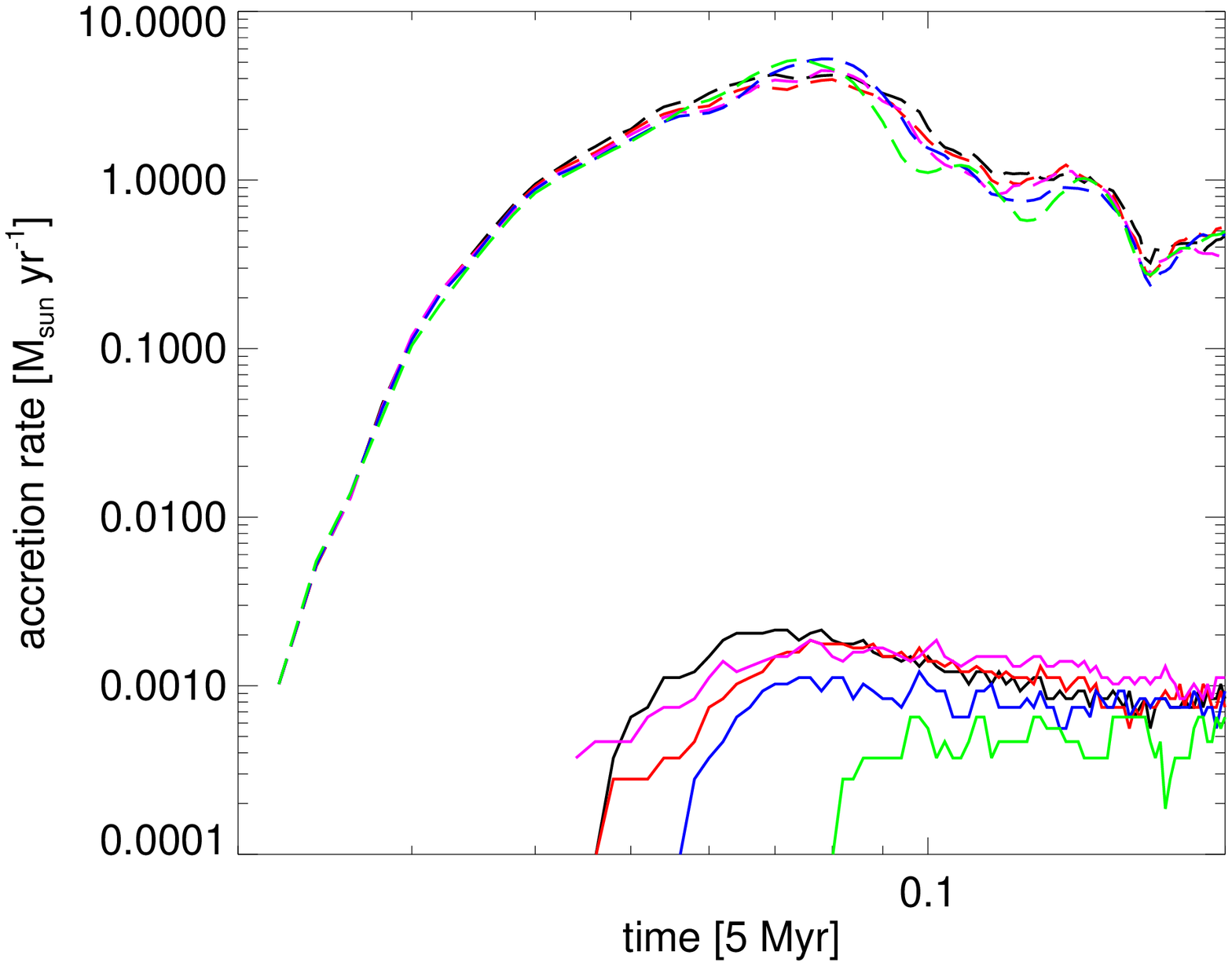,width=0.5\textwidth,angle=0}}
\caption{Accretion rate versus time for convergence tests on the value of the artificial viscosity parameter, $\alpha$. Solid linestyles refer to simulations without turbulence while dashed linestyles refer to simulations with $v_{\rm turb} = 1$. The rotation velocity for all runs here was $v_{\rm rot} = 0.3$. Colors are: $\alpha = 0.1$ (green), $\alpha = 0.3$ (blue), $\alpha = 0.5$ (red), $\alpha = 0.7$ (magenta) and $\alpha = 1.0$ (black; our fiducial value for the simulations in the paper). The simulations with no seeded turbulence show good convergence with a trend towards an increased accretion rate with an increasing $\alpha$, while the simulations with $v_{\rm turb} = 1$ show good convergence but with no clear trend.}
\label{fig:convergence_alpha}
\end{figure}

\begin{figure}
\centerline{\psfig{file=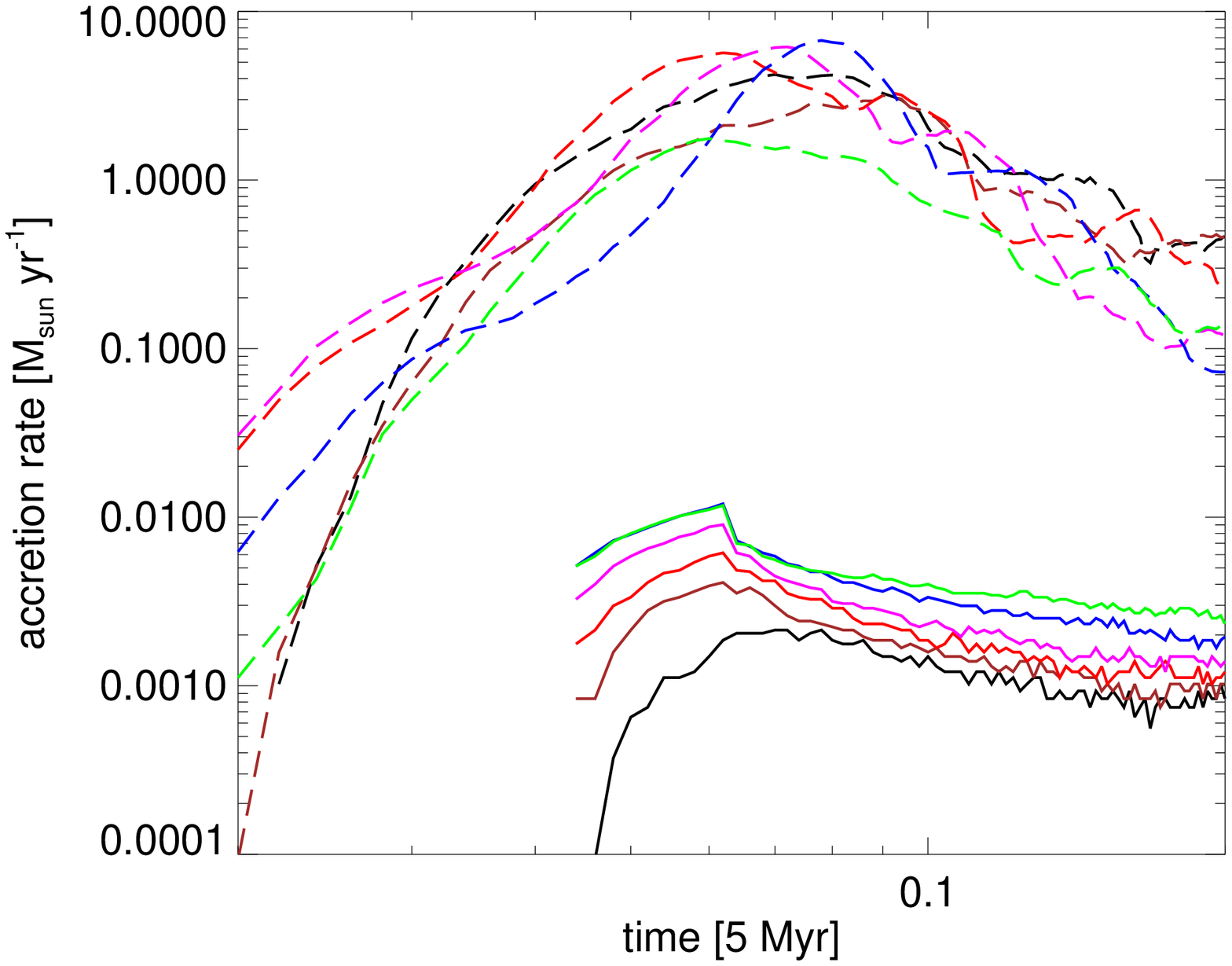,width=0.5\textwidth,angle=0}}
\caption{Accretion rate versus time for convergence tests on the numerical resolution. Once again solid linestyles are for runs with $v_{\rm turb} = 0$ while dashed refer to runs with $v_{\rm turb} = 1$, and all simulations shown used $v_{\rm rot} = 0.3$. Colors refer to $N_{\rm SPH} = 5.4 \times 10^5$ (green), $N_{\rm SPH} = 8.0 \times 10^5$ (blue), $N_{\rm SPH} = 1.2 \times 10^6$ (magenta), $N_{\rm SPH} = 1.8 \times 10^6$ (red), $N_{\rm SPH} = 2.7 \times 10^6$ (brown) and $N_{\rm SPH} = 4 \times 10^6$ (black; our fiducial resolution). In the runs with no seeded turbulence there is a strong trend of decreasing accretion with increasing resolution, which shows signs of starting to converge by the time the highest resolution is reached. In the runs with $v_{\rm turb} = 1$ there is convergence to within a factor of a few but without a clear trend in the accretion rate.}
\label{fig:convergence_resolution}
\end{figure}

A great deal of the accretion that we find in our simulations is the result of turbulent motions creating dense, shocked gas. Due to the artificial viscosity present in the SPH code it is possible that some of this accretion is numerical rather than physical. We have therefore performed convergence tests both on the value of the artificial viscosity parameter, $\alpha$ (cf. Section \ref{sec:numerics}) and the numerical resolution of our fiducial simulations S30 (no turbulence, $v_{\rm rot} = 0.3$) and S35 ($v_{\rm turb} = 1$, $v_{\rm rot} = 0.3$), comparing the accretion rate between the two. The results are shown in Figures \ref{fig:convergence_alpha} and \ref{fig:convergence_resolution} for the artificial viscosity and particle resolution respectively.

For Figure \ref{fig:convergence_alpha} there is clear convergence both in the runs with no turbulence and those with $v_{\rm turb} = 1$. This suggests that the role of the artificial viscosity is negligible in determining the accretion rate through the SMBH accretion radius, at least until the main ballistic mode of feeding ends (it is likely that once the accretion becomes dominated by an accretion disc mode, i.e., to late times, the numerical viscosity will play a greater role).

Figure \ref{fig:convergence_resolution} shows the convergence results on the number of SPH particles. In the runs with no turbulence it is not clear whether the accretion rate has converged with increasing resolution, as we were unable to run simulations with a yet higher number of particles due to computational limitations. However, the plot would seem to suggest that by the highest resolution there is some convergence to late times. It may of course simply be the case that for this particular value of $v_{\rm rot}$ all of the accretion is numerical for the runs with zero turbulence, and the accretion rate will therefore vanish in the limit of increasing resolution. We note that this leaves our conclusions unaffected, and indeed strengthens the case for turbulence enhancing the accretion rate onto the SMBH.  In the runs with $v_{\rm turb} = 1$ there is some reasonable convergence, albeit about a relatively large spread with no clear trend. A possible reason for this is the fact that the SPH method tends to suffer from an inability to converge with increasing resolution in regions where particles are distributed irregularly on the kernel scale (such as in strong shocks), as discussed by \cite{ReadEtal2010}. However, this spread is not large enough to cast any doubt on our main conclusion, and furthermore is only $\simlt$ a factor of a few for the 3 higher resolution runs (red, brown, black lines).

We can therefore be confident that the effect seen in our simulations of an enhanced accretion rate with increasing turbulence (up to saturation) through a ballistic feeding mode is physical rather than numerical.

\end{document}